\documentclass[10pt,journal,compsoc]{IEEEtran}
\usepackage{epsfig}
\pdfoutput=1

\usepackage[bookmarks=false]{hyperref}

\newcommand{\fullonly}[1]{}
\newcommand{\shortonly}[1]{#1}

\newcommand{\video}[1]{}

\newcommand{\tuple}[1]{\langle #1\rangle}
\newcommand{\mean}[1]{\left[ \! \left[ #1 \right]\! \right]}
\newcommand{\setc}[2]{\{#1 \;|\; #2\}}
\newcommand{\intersect}{\cap}

\newcommand{\Op}{{\it Op}}
\newcommand{\uid}{{\rm uid}}
\newcommand{\rid}{{\rm rid}}
\newcommand{\vals}{{\rm Val}_{\rm s}}
\newcommand{\valm}{{\rm Val}_{\rm m}}
\newcommand{\up}{{\it UP}}
\newcommand{\myparagraph}[1]{\paragraph*{\bf #1}}
\newcommand{\attrib}{{\rm attr}}
\newcommand{\attribs}{{\rm attr}_1}
\newcommand{\attribm}{{\rm attr}_{\rm m}}
\newcommand{\union}{\cup}
\newcommand{\UNION}{\bigcup}
\newcommand{\powerset}{{\rm Set}}
\newcommand{\Rho}{{\it Rules}}

\newcommand{\set}[1]{\{#1\}}
\newcommand{\dept}{{\rm dept}}
\newcommand{\cs}{{\rm CS}}

\newcommand{\position}{{\rm position}}
\newcommand{\grad}{{\rm grad}}
\newcommand{\ugrad}{{\rm ugrad}}
\newcommand{\courses}{{\rm courses}}
\newcommand{\type}{{\rm type}}
\newcommand{\uae}{{\rm uae}}
\newcommand{\rae}{{\rm rae}}
\newcommand{\ops}{{\rm ops}}
\newcommand{\con}{{\rm con}}
\newcommand{\uattr}{{\rm uAttr}}
\newcommand{\rattr}{{\rm rAttr}}
\newcommand{\uncovup}{{\it uncovUP}}

\newcommand{\elimOV}{{\it elimOV}}
\newcommand{\cc}{{\it cc}}
\newcommand{\gen}{{\it gen}}
\newcommand{\isempty}{{\rm isEmpty}}
\newcommand{\add}{{\rm add}}
\newcommand{\addall}{{\rm addAll}}
\newcommand{\removeall}{{\rm removeAll}}
\newcommand{\removeElt}{{\rm remove}}
\newcommand{\Qpol}{Q_{\rm pol}}
\newcommand{\toString}{\rm toString}
\newcounter{lnum}
\newcommand{\lnum}{\refstepcounter{lnum}\thelnum:}
\newcommand{\resetlnum}{\setcounter{lnum}{0}}
\newcommand{\uad}{d_{\rm u}}
\newcommand{\rad}{d_{\rm r}}
\newcommand{\Au}{A_{\rm u}}
\newcommand{\Aus}{A_{\rm u,1}}
\newcommand{\Aum}{A_{\rm u,m}}
\newcommand{\Ar}{A_{\rm r}}
\newcommand{\Ars}{A_{\rm r,1}}
\newcommand{\Arm}{A_{\rm r,m}}
\newcommand{\Aunrm}{A_{\rm unrm}}

\newcommand{\aus}{a_{\rm u,1}}
\newcommand{\aum}{a_{\rm u,m}}

\newcommand{\ars}{a_{\rm r,1}}
\newcommand{\arm}{a_{\rm r,m}}
\newcommand{\eu}{e_{\rm u}}
\newcommand{\er}{e_{\rm r}}
\newcommand{\su}{s_{\rm u}}
\newcommand{\so}{s_{\rm o}}
\newcommand{\supseteqin}{\mathrel{\mathord{\supseteq}\mathord{\in}}}
\newcommand{\rhobest}{\rho_{\rm best}}
\newcommand{\rhomerge}{\rho_{\rm merge}}
\newcommand{\wkset}{{\it workSet}}
\newcommand{\stringlit}[1]{\mbox{\tt "}{\rm #1}\mbox{\tt "}}

\newcommand{\ta}{{\it ta}}

\newcommand{\rdtrules}{{\it rdtRules}}

\newcommand{\true}{{\rm true}}
\newcommand{\false}{{\rm false}}
\newcommand{\freq}{{\rm freq}}

\newcommand{\pirbac}{\pi_{\rm RBAC}}
\newcommand{\piabac}{\pi_{\rm ABAC}}
\newcommand{\piacl}{\pi_{\rm ACL}}
\newcommand{\ua}{{\it UA}}
\newcommand{\pa}{{\it PA}}
\newcommand{\pigen}{\pi_{\rm gen}}
\newcommand{\average}{{\rm mean}}
\newcommand{\Nrule}{N_{\rm rule}}
\newcommand{\Ncnj}{N^{\rm min}_{\rm cnj}}
\newcommand{\Ncns}{N^{\rm min}_{\rm cns}}
\newcommand{\Ndept}{N_{\rm dept}}
\newcommand{\Nward}{N_{\rm ward}}
\newcommand{\Pover}{P_{\rm over}}
\newcommand{\com}[1]{{\it #1}}
\newcommand{\meanh}{$\mu$}
\newcommand{\stdv}{$\sigma$}

\newcommand{\avgUPrule}{\widehat{|\!\mean{\rho}\!|}}
\newcommand{\botfrac}{\nu_\bot}

\newcommand{\function}{{\bf function}}
\newcommand{\ifstmt}{{\bf if}}
\newcommand{\elsestmt}{{\bf else}}
\newcommand{\eifstmt}{{\bf end if}}
\newcommand{\forloop}{{\bf for}}
\newcommand{\whileloop}{{\bf while}}
\newcommand{\eforloop}{{\bf end for}}
\newcommand{\ewhileloop}{{\bf end while}}
\newcommand{\forloopin}{{\bf in}}
\newcommand{\forloopto}{{\bf to}}
\newcommand{\return}{{\bf return}}

\newcommand{\computeuae}{{\rm computeUAE}}
\newcommand{\computerae}{{\rm computeRAE}}
\newcommand{\addcandidaterule}{{\rm addCandidateRule}}
\newcommand{\candidateconstraint}{{\rm candidateConstraint}}
\newcommand{\elimattrib}{{\rm elimAttribute}}
\newcommand{\elimconstraints}{{\rm elimConstraints}}

\newcommand{\elimelements}{{\rm elimElements}}
\newcommand{\elimconjuncts}{{\rm elimConjuncts}}
\newcommand{\elimconjunctshelper}{{\rm elimConjunctsHelper}}
\newcommand{\elimoverlapVal}{{\rm elimOverlapVal}}
\newcommand{\elimoverlapOp}{{\rm elimOverlapOp}}
\newcommand{\elimredundantsets}{{\rm elimRedundantSets}}
\newcommand{\generalizerule}{{\rm generalizeRule}}
\newcommand{\maxConjunctSz}{{\rm maxConjunctSz}}
\newcommand{\mergerules}{{\rm mergeRules}}
\newcommand{\Qup}{Q_{\rm up}}
\newcommand{\Qrul}{Q_{\rm rul}}
\newcommand{\simplifyrules}{{\rm simplifyRules}}
\newcommand{\wsc}{{\rm WSC}}

\newcommand{\addcandidateruleL}{\hyperlink{addcandidaterule}{\addcandidaterule}}
\newcommand{\computeuaeL}{\hyperlink{computeuae}{\computeuae}}
\newcommand{\computeraeL}{\hyperlink{computerae}{\computerae}}
\newcommand{\candidateconstraintL}{\hyperlink{candidateconstraint}{\candidateconstraint}}
\newcommand{\elimattribL}{\hyperlink{elimattrib}{\elimattrib}}
\newcommand{\elimconstraintsL}{\hyperlink{elimconstraints}{\elimconstraints}}

\newcommand{\elimelementsL}{\hyperlink{elimelements}{\elimelements}}
\newcommand{\elimconjunctsL}{\hyperlink{elimconjuncts}{\elimconjuncts}}
\newcommand{\elimconjunctshelperL}{\hyperlink{elimconjunctshelper}{\elimconjunctshelper}}
\newcommand{\elimoverlapValL}{\hyperlink{elimoverlapVal}{\elimoverlapVal}}
\newcommand{\elimoverlapOpL}{\hyperlink{elimoverlapOp}{\elimoverlapOp}}
\newcommand{\elimredundantsetsL}{\hyperlink{elimredundantsets}{\elimredundantsets}}
\newcommand{\generalizeruleL}{\hyperlink{generalizerule}{\generalizerule}}
\newcommand{\maxConjunctSzL}{\hyperlink{maxConjunctSz}{\maxConjunctSz}}
\newcommand{\mergerulesL}{\hyperlink{mergerules}{\mergerules}}
\newcommand{\QupL}{\hyperlink{Qup}{\Qup}}
\newcommand{\QrulL}{\hyperlink{Qrul}{\Qrul}}
\newcommand{\simplifyrulesL}{\hyperlink{simplifyrules}{\simplifyrules}}
\newcommand{\wscPolL}{\hyperlink{wscPol}{\wsc}}
\newcommand{\wscRuleL}{\hyperlink{wscRule}{\wsc}}
\newcommand{\wscRulesL}{\hyperlink{wscRules}{\wsc}}

\newcommand{\Copy}{{\rm copy}}

\newcommand{\lc}{{\it lc}_{\rm u,m}}
\newcommand{\lm}{{\it lm}}
\newcommand{\lexpr}{{\it le}}

\newcommand{\synSim}{{\rm ss}}
\newcommand{\uaeSim}{\synSim_{\rm u}}
\newcommand{\raeSim}{\synSim_{\rm r}}
\newcommand{\rules}{{\rm rules}}

\begin{document}

\title{Mining Attribute-based Access Control Policies\vspace*{-1ex}
\thanks{This material is based upon work supported in part by %
ONR under Grant N00014-07-1-0928 
and NSF under Grants CNS-0831298 
and CNS-1421893. 
Submitted to IEEE TDSC.  \copyright 2014 IEEE. Personal use of this material is permitted. Permission from IEEE must be obtained for all other uses, in any current or future media, including reprinting/republishing this material for advertising or promotional purposes, creating new collective works, for resale or redistribution to servers or lists, or reuse of any copyrighted component of this work in other works.
}}

\author{Zhongyuan Xu and Scott D. Stoller\\
  Computer Science Department, Stony Brook University}



\maketitle

\begin{abstract}
  Attribute-based access control (ABAC) provides a high level of
  flexibility that promotes security and information sharing.  ABAC policy
  mining algorithms have potential to significantly reduce the cost of
  migration to ABAC, by partially automating the development of an ABAC
  policy from an access control list (ACL) policy or role-based access
  control (RBAC) policy with accompanying attribute data.  This paper
  presents an ABAC policy mining algorithm. To the best of our knowledge,
  it is the first ABAC policy mining algorithm.  Our algorithm iterates
  over tuples in the given user-permission relation, uses selected tuples
  as seeds for constructing candidate rules, and attempts to generalize
  each candidate rule to cover additional tuples in the user-permission
  relation by replacing conjuncts in attribute expressions with
  constraints.  Our algorithm attempts to improve the policy by merging and
  simplifying candidate rules, and then it selects the highest-quality
  candidate rules for inclusion in the generated policy.
\end{abstract}



\section{Introduction}
\label{sec:intro}

Attribute-based access control (ABAC) provides a high level of flexibility
that promotes security and information sharing \cite{NIST13ABAC}.  ABAC
also overcomes some of the problems associated with RBAC
\cite{sandhu96role}, notably role explosion
\cite{NIST13ABAC,nextlabs13managing}.  The benefits of ABAC led the Federal
Chief Information Officer Council to call out ABAC as a recommended access
control model in the Federal Identity Credential and Access Management
Roadmap and Implementation Guidance, ver.\ 2.0 \cite{NIST13ABAC,FEDCIO11}.


Manual development of RBAC policies can be time-consuming and expensive
\cite{hachana12role}.  Role mining algorithms promise to drastically reduce
the cost, by partially automating the development of RBAC policies
\cite{hachana12role}.  Role mining is an active research area and a
currently relatively small (about \$70 million) but rapidly growing
commercial market segment \cite{hachana12role}.  Similarly, manual
development of ABAC policies can be difficult \cite{beckerle13formal} and
expensive \cite{NIST13ABAC}.  ABAC policy mining algorithms have 
potential to reduce the cost of ABAC policy development.



The main contribution of this paper is an algorithm for ABAC policy mining.
Our algorithm is formulated to mine an ABAC policy from ACLs and attribute
data.  It can be used to mine an ABAC policy from an RBAC policy and
attribute data, by expanding the RBAC policy into ACLs, adding a ``role''
attribute to the attribute data (to avoid information loss), and then
applying our algorithm.  At a high level, our algorithm works as follows.
It iterates over tuples in the given user-permission relation, uses
selected tuples as seeds for constructing candidate rules, and attempts to
generalize each candidate rule to cover additional tuples in the
user-permission relation by replacing conjuncts in attribute expressions
with constraints.  After constructing candidate rules that together cover
the entire user-permission relation, it attempts to improve the policy by
merging and simplifying candidate rules.  Finally, it selects the
highest-quality candidate rules for inclusion in the generated policy.  We
also developed an extension of the algorithm to identify suspected noise in
the input. 

Section \ref{sec:evaluation} presents results from evaluating the algorithm
on some relatively small but non-trivial hand-written sample policies and
on synthetic (i.e., pseudorandomly generated) policies.  The general
methodology is to start with an ABAC policy (including attribute data),
generate an equivalent ACL policy from the ABAC policy, add noise (in some
experiments) to the ACL policy and attribute data, run our algorithm on the
resulting ACL policies and attribute data, and compare the mined ABAC
policy with the original ABAC policy.





\section{ABAC policy language}
\label{sec:language}

This section presents our ABAC policy language.  We do not consider policy
administration, since our goal is to mine a single ABAC policy from the
current low-level policy.  We present a specific concrete policy language,
rather than a flexible framework, to simplify the exposition and evaluation
of our policy mining algorithm, although our approach is general and can be
adapted to other ABAC policy languages.  Our ABAC policy language contains
all of the common ABAC policy language constructs, except arithmetic
inequalities and negation.  Extending our algorithm to handle those
constructs is future work.  The policy language handled in this paper is
already significantly more complex than policy languages handled in
previous work on security policy mining.

ABAC policies refer to attributes of users and resources.  Given a set $U$
of users and a set $\Au$ of user attributes, user attribute data is
represented by a function $\uad$ such that $\uad(u,a)$ is the value of
attribute $a$ for user $u$.  There is a distinguished user attribute $\uid$
that has a unique value for each user.  Similarly, given a set $R$ of
resources and a set $\Ar$ of resource attributes, resource attribute data
is represented by a function $\rad$ such that $\rad(r,a)$ is the value of
attribute $a$ for resource $r$.  There is a distinguished resource
attribute $\rid$ that has a unique value for each resource.  We assume the
set $\Au$ of user attributes can be partitioned into a set $\Aus$ of {\it
  single-valued user attributes} which have atomic values, and a set $\Aum$
of {\it multi-valued user attributes} whose values are sets of atomic
values.  Similarly, we assume the set $\Ar$ of resource attributes can be
partitioned into a set $\Ars$ of {\it single-valued resource attributes}
and a set of $\Arm$ of {\it multi-valued resource attributes}.  Let $\vals$
be the set of possible atomic values of attributes.  We assume $\vals$
includes a distinguished value $\bot$ used to indicate that an attribute's
value is unknown.  The set of possible values of multi-valued attributes is
$\valm=\powerset(\vals\setminus\set{\bot})\union\bot$, where $\powerset(S)$
is the powerset of set $S$.

Attribute expressions are used to express the sets of users and resources
to which a rule applies.  A {\em user-attribute expression} (UAE) is a
function $e$ such that, for each user attribute $a$, $e(a)$ is either the
special value $\top$, indicating that $e$ imposes no constraint on the
value of attribute $a$, or a set (interpreted as a disjunction) of possible
values of $a$ excluding $\bot$ (in other words, a subset of
$\vals\setminus\set{\bot}$ or $\valm\setminus\set{\bot}$, depending on
whether $a$ is single-valued or multi-valued).  We refer to the set $e(a)$
as the {\em conjunct} for attribute $a$.  We say that expression $e$ {\em
  uses} an attribute $a$ if $e(a)\ne \top$.  Let $\attrib(e)$ denote the
set of attributes used by $e$.  Let $\attribs(e)$ and $\attribm(e)$ denote
the sets of single-valued and multi-valued attributes, respectively, used
by $e$.


A user $u$ {\em satisfies} a user-attribute expression $e$, denoted $u
\models e$, iff $(\forall a\in \Aus.\, e(a)=\top \lor \exists v\in e(a).\,
\uad(u,a)=v)$ and $(\forall a\in \Aum.\, e(a)=\top \lor \exists v\in
e(a).\, \uad(u,a)\supseteq v)$.  For multi-valued attributes, we use the
condition $\uad(u,a)\supseteq v$ instead of $\uad(u,a)=v$ because elements
of a multi-valued user attribute typically represent some type of
capabilities of a user, so using $\supseteq$ expresses that the user has
the specified capabilities and possibly more.

For example, suppose $\Aus =\{\dept, \position\}$ and $\Aum =\{\courses\}$.
The function $e_1$ with $e_1(\dept) = \{\cs\}$ and $e_1(\position) =
\{\grad,\ugrad\}$ and $e_1(\courses)=\set{\set{{\rm CS101}, {\rm CS102}}}$
is a user-attribute expression satisfied by users in the $\cs$ department
who are either graduate or undergraduate students and whose courses include
CS101 and CS102 (and possibly other courses).

We introduce a concrete syntax for attribute expressions, for improved
readability in examples.  We write a user attribute expression as a
conjunction of the conjuncts not equal to $\top$.  Suppose $e(a)\ne\top$.
Let $v=e(a)$.  When $a$ is single-valued, we write the conjunct for $a$ as
$a \in v$; as syntactic sugar, if $v$ is a singleton set $\set{s}$, we may
write the conjunct as $a=s$.  When $a$ is multi-valued, we write the
conjunct for $a$ as $a \supseteqin v$ (indicating that $a$ is a superset of
an element of $v$); as syntactic sugar, if $v$ is a singleton set
$\set{s}$, we may write the conjunct as $a\supseteq s$.  For example, the
above expression $e_1$ may be written as $\dept=\cs \land \position \in
\set{\ugrad,\grad} \land \courses \supseteq \set{{\rm CS101}, {\rm
    CS102}}$.  For an example that uses $\supseteqin$, the expression $e_2$
that is the same as $e_1$ except with $e_2(\courses)=\set{\set{{\rm
      CS101}}, \set{{\rm CS102}}}$ may be written as $\dept=\cs \land
\position \in \set{\ugrad,\grad} \land \courses \supseteqin \set{\set{{\rm
      CS101}}, \set{{\rm CS102}}}$, and is satisfied by graduate or
undergraduate students in the $\cs$ department whose courses include either
CS101 or CS102.

The {\em meaning} of a user-attribute expression $e$, denoted $\mean{e}_U$,
is the set of users in $U$ that satisfy it: $\mean{e}_U = \setc{u \in U}{u
  \models e}$.  User attribute data is an implicit argument to
$\mean{e}_U$.  We say that $e$ {\em characterizes} the set $\mean{e}_U$.

%
%

A {\em resource-attribute expression} (RAE) is defined similarly, except
using the set $\Ar$ of resource attributes instead of the set $\Au$ of user
attributes.  The semantics of RAEs is defined similarly to the semantics of
UAEs, except simply using equality, not $\supseteq$, in the condition for
multi-valued attributes in the definition of ``satisfies'', because we do
not interpret elements of multi-valued resource attributes specially (e.g.,
as capabilities).

In ABAC policy rules, constraints are used to express relationships between
users and resources.  An {\em atomic constraint} is a formula $f$ of the
form $\aum\supseteq\arm$, $\aum\ni\ars$, or $\aus=\ars$, where
$\aus\in\Aus$, $\aum\in\Aum$, $\ars\in\Ars$, and $\arm\in\Arm$.  The first
two forms express that user attributes contain specified values.  This is a
common type of constraint, because user attributes typically represent some
type of capabilities of a user.  Other forms of atomic constraint are
possible (e.g., $\aum\subseteq\arm$) but less common, so we leave them for
future work.  Let $\uattr(f)$ and $\rattr(f)$ refer to the user attribute
and resource attribute, respectively, used in $f$.  User $u$ and resource
$r$ {\em satisfy} an atomic constraint $f$, denoted $\tuple{u, r} \models
f$, if $\uad(u,\uattr(f))\ne\bot$ and $\rad(u,\rattr(f))\ne\bot$ and
formula $f$ holds when the values $\uad(u,\uattr(f))$ and
$\rad(u,\rattr(f))$ are substituted in it.

A {\em constraint} is a set (interpreted as a conjunction) of atomic
constraints.  User $u$ and resource $r$ {\em satisfy} a constraint $c$,
denoted $\tuple{u, r} \models c$, if they satisfy every atomic constraint
in $c$.  In examples, we write constraints as conjunctions instead of sets.
For example, the constraint ``{\rm specialties} $\supseteq$ {\rm topics}
$\,\land\,$ {\rm teams} $\ni$ {\rm treatingTeam}'' is satisfied by user $u$ and
resource $r$ if the user's specialties include all of the topics associated
with the resource, and the set of teams associated with the user contains
the treatingTeam associated with the resource.

A {\em user-permission tuple} is a tuple $\tuple{u, r, o}$ containing a
user, a resource, and an operation.  This tuple means that user $u$ has
permission to perform operation $o$ on resource $r$.  A {\em
  user-permission relation} is a set of such tuples.

A {\em rule} is a tuple $\tuple{\eu, \er, O, c}$, where $\eu$ is a
user-attribute expression, $\er$ is a resource-attribute expression, $O$ is
a set of operations, and $c$ is a constraint.  For a rule $\rho=\tuple{\eu,
  \er, O, c}$, let $\uae(\rho)=\eu$, $\rae(\rho)=\er$, $\ops(\rho)=O$, and
$\con(\rho)=c$.  For example, the rule $\langle$true, type=task $\land$
proprietary=false, \{read, request\}, projects $\ni$ project $\land$
expertise $\supseteq$ expertise$\rangle$ used in our project management
case study can be interpreted as ``A user working on a project can read and
request to work on a non-proprietary task whose required areas of expertise
are among his/her areas of expertise.''  User $u$, resource $r$, and
operation $o$ {\em satisfy} a rule $\rho$, denoted $\tuple{u, r, o} \models
\rho$, if $u\models \uae(\rho) \land r\models\rae(\rho) \land o \in
\ops(\rho) \land \tuple{u,r}\models \con(\rho)$.


An {\em ABAC policy} is a tuple $\tuple{U, R, \Op, \Au, \Ar, \uad, \rad,
  \Rho}$, where $U$, $R$, $\Au$, $\Ar$, $\uad$, and $\rad$ are as described
above, $\Op$ is a set of operations, and $\Rho$ is a set of rules.

The user-permission relation induced by a rule $\rho$ is $\mean{\rho} =
\setc{\tuple{u,r,o} \in U\times R\times \Op}{\tuple{u,r,o}\models \rho}$.
Note that $U$, $R$, $\uad$, and $\rad$ are implicit arguments to
$\mean{\rho}$.

The user-permission relation induced by a policy $\pi$ with the above form
is $\mean{\pi} = \UNION_{\rho \in \Rho} \mean{\rho}$.



\section{The ABAC Policy Mining Problem}
\label{sec:problem}

An {\em access control list (ACL) policy} is a tuple $\tuple{U, R, \Op,
  \up_0}$, where $U$ is a set of users, $R$ is a set of resources, $\Op$ is
a set of operations, and $\up_0\subseteq U\times R\times \Op$ is a
user-permission relation, obtained from the union of the access control
lists.

An ABAC policy $\pi$ is {\em consistent} with an ACL policy $\tuple{U,
  P,\linebreak[0] \Op, \up_0}$ if they have the same sets of users,
resource, and operations and $\mean{\pi} = \up_0$.

An ABAC policy consistent with a given ACL policy can be trivially
constructed, by creating a separate rule corresponding to each
user-permission tuple in the ACL policy, simply using $\uid$ and $\rid$ to
identify the relevant user and resource.  Of course, such an ABAC policy is
as verbose and hard to manage as the original ACL policy.  This observation
forces us to ask: among ABAC policies semantically consistent with a given
ACL policy $\pi_0$, which ones are preferable?  We adopt two criteria.

One criterion is that policies that do not use the attributes $\uid$ and
$\rid$ are preferable, because policies that use $\uid$ and $\rid$ are
partly identity-based, not entirely attribute-based.  Therefore, our
definition of ABAC policy mining requires that these attributes are used
only if necessary, i.e., only if every ABAC policy semantically consistent
with $\pi_0$ contains rules that use them.

The other criterion is to maximize a policy quality metric.  A {\em policy
  quality metric} is a function $\Qpol$ from ABAC policies to a
totally-ordered set, such as the natural numbers.  The ordering is chosen
so that small values indicate high quality; this is natural for metrics
based on policy size.  For generality, we parameterize the policy mining
problem by the policy quality metric.

The {\em ABAC policy mining problem} is: given an ACL policy
$\pi_0=\tuple{U, R, \Op, \up_0}$, user attributes $\Au$, resource
attributes $\Ar$, user attribute data $\uad$, resource attribute data
$\rad$, and a policy quality metric $\Qpol$, find a set $\Rho$ of rules
such that the ABAC policy $\pi=\tuple{U, R, \Op, \Au, \Ar, \uad, \rad,
  \Rho}$ that (1) is consistent with $\pi_0$, (2) uses $\uid$ only when
necessary, (3) uses $\rid$ only when necessary, and (4) has the best
quality, according to $\Qpol$, among such policies.

The policy quality metric that our algorithm aims to optimize is {\em
  weighted structural complexity} (WSC) \cite{molloy10mining}, a
generalization of policy size.  This is consistent with usability studies
of access control rules, which conclude that more concise policies are more
manageable \cite{beckerle13formal}.  Informally, the WSC of an ABAC policy
is a weighted sum of the number of elements in the policy.  Formally, the
WSC of an ABAC policy $\pi$ with rules $\Rho$ is
$\hypertarget{wscPol}{\wsc(\pi)}=\wscRulesL(\Rho)$, defined by
\begin{eqnarray*}
  \wsc(e) &=&\!\!\! \sum_{a \in \attribs(e)} \!|e(a)| + \sum_{a \in \attribm(e), s \in e(a)} \!|s|\\
  \hypertarget{wscRule}{\wsc(\tuple{\eu, \er, O, c})} &=&
  \begin{array}[t]{@{}l@{}}
    w_1\wsc(\eu) + w_2\wsc(\er)\smallskip\\
    {} + w_3|O| + w_4|c|
  \end{array}\\
  \hypertarget{wscRules}{\wsc(\Rho)} &=& \sum_{\rho \in \Rho} \wsc(\rho),
\end{eqnarray*}
where $|s|$ is the cardinality of set $s$, and the $w_i$ are user-specified
weights.

\myparagraph{Computational Complexity}

We show that the ABAC policy mining problem is NP-hard, by reducing the
Edge Role Mining Problem (Edge RMP) \cite{lu08optimal} to it.  NP-hardness
of Edge RMP follows from Theorem 1 in \cite{molloy10mining}.  The basic
idea of the reduction is that an Edge RMP instance $I_R$ is translated into
an ABAC policy mining problem instance $I_A$ with $\uid$ and $\rid$ as the
only attributes.  Given a solution $\piabac$ to problem instance $I_A$, the
solution to $I_R$ is constructed by interpreting each rule as a role.
Details of the reduction appear in Section \ref{sec:proof-NP-hard} in the
Supplemental Material.

It is easy to show that a decision-problem version of ABAC policy mining
is in NP.  The decision-problem version asks whether there exists an ABAC
policy that meets conditions (1)--(3) in the above definition of the ABAC
policy mining problem and has WSC less than or equal to a given value.


\section{Policy Mining Algorithm}
\label{sec:algorithm}

Top-level pseudocode for our policy mining algorithm appears in Figure
\ref{fig:alg}.  It reflects the high-level structure described in Section
\ref{sec:intro}.  Functions called by the top-level pseudocode are
described next.  Function names hyperlink to pseudocode for the function,
if it is included in the paper, otherwise to a description of the function.
An example illustrating the processing of a user-permission tuple by our
algorithm appears in Section \ref{sec:working-example} in the Supplemental
Material.  For efficiency, our algorithm incorporates heuristics and is not
guaranteed to generate a policy with minimal WSC.



The function $\addcandidateruleL(\su, s_r, \so, \cc, \uncovup,$\\
$\Rho)$ in Figure \ref{fig:addcandidaterule} first calls $\computeuaeL$ to
compute a user-attribute expression $\eu$ that characterizes $\su$, then
calls and $\computeraeL$ to compute a resource-attribute expression $\er$
that characterizes $s_r$. It then calls $\generalizeruleL(\rho, \cc,
\uncovup, \Rho)$ to generalize the rule $\rho=\tuple{\eu, \er, \so,
  \emptyset}$ to $\rho'$ and adds $\rho'$ to candidate rule set $\Rho$. The
details of the functions called by $\addcandidaterule$ are described next.

The function
\shortonly{\hypertarget{computeuae}{$\computeuaeL(s,U)$}}\fullonly{$\computeuaeL(s,U)$
  in Figure \ref{fig:computeuae}} computes a user-attribute expression
$\eu$ that characterizes the set $s$ of users.  The conjunct for each
attribute $a$ contains the values of $a$ for users in $s$, unless one of
those values is $\bot$, in which case $a$ is unused (i.e., the conjunct for
$a$ is $\top$).  Furthermore, the conjunct for $\uid$ is removed if the
resulting attribute expression still characterizes $s$; this step is useful
because policies that are not identity-based generalize better.\shortonly{
  Similarly, \hypertarget{computerae}{$\computerae(s,R)$} computes a
  resource-attribute expression that characterizes the set $s$ of
  resources.  }\fullonly{ After constructing a candidate expression $e$, it
  calls \hypertarget{elimredundantsets}{$\elimredundantsets(e)$}, which
  attempts to lower the WSC of $e$ by examining the conjunct for each
  multi-valued user attribute, and removing each set that is a superset of
  another set in the same conjunct; this leaves the meaning of the rule
  unchanged, because $\supseteq$ is used in the condition for multi-valued
  attributes in the semantics of user attribute expressions.
Pseudocode for $\elimredundantsetsL$ is straightforward and omitted.

The function \hypertarget{computerae}{$\computerae(s,R)$} computes a
resource-attribute expression that characterizes the set $s$ of resources.
The definition is the same as for $\computeuaeL$, except using resource
attributes instead of user attributes, and the call to
$\elimredundantsetsL$ is omitted.  Pseudocode for $\computeraeL$ is
omitted.

} The attribute expressions returned by $\computeuaeL$ and $\computeraeL$
might not be minimum-sized among expressions that characterize $s$: it is
possible that some conjuncts can be removed.  
We defer minimization of the attribute expressions until after the call to
$\generalizeruleL$ (described below), because minimizing them before that
would reduce opportunities to find relations between values of user
attributes and resource attributes in $\generalizeruleL$.

The function \hypertarget{candidateconstraint}{$\candidateconstraint(r,u)$}
returns a set containing all the atomic constraints that hold between
resource $r$ and user $u$. Pseudocode for $\candidateconstraint$ is
straightforward and omitted.

A rule $\rho'$ is {\em valid} if $\mean{\rho'}\subseteq\up_0$.

The function $\generalizeruleL(\rho, \cc, \uncovup, \Rho)$ in Figure
\ref{fig:generalizerule} attempts to generalize rule $\rho$ by adding some
of the atomic constraints $f$ in $\cc$ to $\rho$ and eliminating the
conjuncts of the user attribute expression and the resource attribute
expression corresponding to the attributes used in $f$, i.e., mapping those
attributes to $\top$.  If the resulting rule is invalid, the function
attempts a more conservative generalization by eliminating only one of
those conjuncts, keeping the other.  We call a rule obtained in this way a
{\em generalization} of $\rho$.  Such a rule is more general than $\rho$ in
the sense that it refers to relationships instead of specific values.
Also, the user-permission relation induced by a generalization of $\rho$ is
a superset of the user-permission relation induced by $\rho$.

If there are no valid generalizations of $\rho$, $\generalizeruleL(\rho,
\cc, \uncovup, \Rho)$ returns $\rho$.  If there is a valid generalization
of $\rho$, $\generalizeruleL(\rho, \cc,$ $\uncovup, \Rho)$ returns the
generalization $\rho'$ of $\rho$ with the best quality according to a given
rule quality metric.  Note that $\rho'$ may cover tuples that are already
covered (i.e., are in $\up$); in other words, our algorithm can generate
policies containing rules whose meanings overlap.  A {\em rule quality
  metric} is a function $\Qrul(\rho, \up)$ that maps a rule $\rho$ to a
totally-ordered set, with the ordering chosen so that larger values
indicate high quality.  The second argument $\up$ is a set of
user-permission tuples.
Based on our primary goal of minimizing the generated policy's WSC, and a
secondary preference for rules with more constraints, we define
\begin{displaymath}
  \hypertarget{Qrul}{\Qrul(\rho, \up)} = \tuple{|\mean{\rho}\intersect\up|/\wscRuleL(\rho),\, |\con(\rho)|}.
\end{displaymath}
The secondary preference for more constraints is a heuristic, based on the
observation that rules with more constraints tend to be more general than
other rules with the same $|\mean{\rho}\intersect\up|/\wscRuleL(\rho)$
(such rules typically have more conjuncts) and hence lead to lower WSC.  In
$\generalizeruleL$, $\uncovup$ is the second argument to $\QrulL$, so
$\mean{\rho}\intersect\up$ is the set of user-permission tuples in $\up_0$
that are covered by $\rho$ and not covered by rules already in the policy.
The loop over $i$ near the end of the pseudocode for $\generalizeruleL$
considers all possibilities for the first atomic constraint in $\cc$ that
gets added to the constraint of $\rho$.  The function calls itself
recursively to determine the subsequent atomic constraints in $c$ that get
added to the constraint.

%


\begin{figure}[tbp]
\resetlnum
\begin{tabbing}
~~~~~\= // \com{$\Rho$ is the set of candidate rules}\\
\lnum\label{tl:init-Rho}\> $\Rho=\emptyset$\\
\>     // \com{$\uncovup$ contains user-permission tuples in $\up_0$}\\
\>     // \com{that are not covered by $\Rho$}\\
\lnum\label{tl:init-uncov}\> $\uncovup=\up_0.\Copy()$\\
\lnum\label{tl:while-uncov}\> \whileloop\ $\neg\uncovup.\isempty()$\\
\>~~~~\=    // \com{Select an uncovered user-permission tuple.}\\
\lnum\label{tl:select-uncov}\>\>   $\tuple{u,r,o}$ = some tuple in $\uncovup$\\
\lnum\label{tl:cc}\>\>   $\cc = \candidateconstraintL(r, u)$\\
\>\>       // \com{$\su$ contains users with permission $\tuple{r,o}$ and}\\
\>\>       // \com{that have the same candidate constraint for $r$ as $u$}\\
\lnum\label{tl:s}\>\>   $\su = \{u'\in U \;|$ \= $\tuple{u',r,o}\in \up_0$\\
\lnum\>\>\>             ${} \land \candidateconstraintL(r,u')=\cc\}$\\
\lnum\label{tl:addcandidaterule1}\>\> $\addcandidaterule(\su, \set{r}, \set{o}, \cc, \uncovup, \Rho)$\\
\>\> // \com{$\so$ is set of operations that $u$ can apply to $r$}\\
\lnum\>\>   $\so = \setc{o'\in \Op}{\tuple{u,r,o'}\in \up_0}$\\
\lnum\label{tl:addcandidaterule2}\>\> $\addcandidaterule(\set{u}, \set{r}, \so, \cc, \uncovup, \Rho)$\\
\lnum\label{tl:endwhile-uncov}\> \ewhileloop\\
\>      // \com{Repeatedly merge and simplify rules, until}\\
\>      // \com{this has no effect}\\
\lnum\label{tl:mergerules-1}\> $\mergerulesL(\Rho)$\\
\lnum\label{tl:simplifyrules}\> \whileloop\ $\simplifyrulesL(\Rho) \mathrel{\&\&} \mergerulesL(\Rho)$\\
\lnum\>\> skip\\
\lnum\label{tl:endwhile-simplifyrules}\> \ewhileloop\\
\> // \com{Select high quality rules into final result $\Rho'$.}\\
\lnum\> $\Rho'$ = $\emptyset$\\
\lnum\> Repeatedly select the highest quality rules from\\
\> $\Rho$ to $\Rho'$ until $\sum_{\rho\in\Rho'}\mean{\rho} = \up_0$,\\
\> using $\up_0\setminus\mean{\Rho'}$ as second argument to $\Qrul$\\
\lnum\> \return\ $\Rho'$
\end{tabbing}
\caption{Policy mining algorithm.}
\label{fig:alg}
\end{figure}

\begin{figure}[tbp]
\resetlnum
\begin{tabbing}
  ~~~~~\= \hypertarget{addcandidaterule}{\function\ $\addcandidaterule(\su, s_r, \so, \cc, \uncovup, \Rho)$} \\
\>    // \com{Construct a rule $\rho$ that covers user-permission }\\
\>    // \com{tuples $\setc{\tuple{u,r,o}}{u\in \su\land r\in s_r \land o\in \so}$.}\\
\lnum\label{ccr:computeuae}\>   $\eu = \computeuaeL(\su,U)$\\
\lnum\label{ccr:computerae}\>   $\er = \computeraeL(s_r,R)$\\
\lnum\label{ccr:r}\>   $\rho = \tuple{\eu, \er, \so, \emptyset}$\\
\lnum\label{ccr:generalizerule}\>   $\rho' = \generalizeruleL(\rho, \cc, \uncovup, \Rho)$\\
\lnum\label{ccr:R-add}\>   $\Rho.\add(\rho')$\\
\lnum\label{ccr:uncovup-removeall}\>   $\uncovup.\removeall(\mean{\rho'})$
\end{tabbing}
\caption{Compute a candidate rule $\rho'$ and add $\rho'$ to candidate rule set $\Rho$}
\label{fig:addcandidaterule}
\end{figure}

\fullonly{
\begin{figure}[tbp]
\resetlnum
\begin{tabbing}
~~~~~\= \hypertarget{computeuae}{\function\ $\computeuae(s,U)$}\\
\> // \com{Try to characterize $s$ without using $\uid$.  Use all}\\
\> // \com{other attributes which have known values for all}\\
\> // \com{users in $s$.}\\
\lnum\> $e = (\lambda\, a \in \Au.\, $\\
\lnum\>~~~  $a\!=\!\uid \,\lor\, (\exists\, u\in s.\, \uad(u,a)=\bot)
         \mathop{?} \top
         : \UNION_{u\in s} \uad(u,a))$\\
\lnum\>  \ifstmt\ $\mean{e}_U \ne s$\\
\lnum\>~~~~\= // \com{$\uid$ is needed to characterize $s$}\\
\>\>          $e(\uid)=\UNION_{u\in s} \uad(u,\uid)$\\
\lnum\> \eifstmt\\
\lnum\> $\elimredundantsets(e)$\\
\lnum\> \return\ $e$
\end{tabbing}
\caption{Compute a user-attribute expression that characterizes set $s$ of
  users, where $U$ is the set of all users. }
\label{fig:computeuae}
\end{figure}
}

\begin{figure}[tbp]
\resetlnum
\begin{tabbing}
~~~~~\= \hypertarget{generalizerule}{\function\ $\generalizerule(\rho, \cc, \uncovup, \Rho)$}\\
     \> // \com{$\rhobest$ is highest-quality generalization of $\rho$}\\
\lnum\> $\rhobest = \rho$\\
     \> // \com{$\cc'$ contains formulas from $\cc$ that lead to valid}\\
     \> // \com{generalizations of $\rho$.}\\
\lnum\> $\cc' = \mbox{new Vector()}$\\
\lnum\> // \com{$\gen[i]$ is a generalization of $\rho$ using $\cc'[i]$}\\
\lnum\> $\gen = \mbox{new Vector()}$\\
\>      // \com{find formulas in $\cc$ that lead to valid}\\
\>      // \com{generalizations of $\rho$.}\\
\lnum\> \forloop\ $f$ \forloopin\ $\cc$\\
     \>~~~~\= // \com{try to generalize $\rho$ by adding $f$ and elimi-}\\
     \>\>     // \com{nating conjuncts for both attributes used in $f$.}\\
\lnum\>\>     $\rho' = \langle$\=$\uae(\rho)[\uattr(f)\mapsto\top],
  \rae(\rho)[\rattr(f)\mapsto\top], $\\
\lnum\>\>\> $\ops(\rho), \con(\rho)\union\set{f}\rangle$\\
  \>~~~~\= // \com{check if $\mean{\rho'}$ is a valid rule}\\
\lnum\>\> \ifstmt\ $\mean{\rho'} \subseteq \up_0$\\
\lnum\>\>~~~~\= $\cc'.add(f)$\\
\lnum\>\>\>     $\gen.add(\rho')$\\
\lnum\>\> \elsestmt\\
      \>\>\> // \com{try to generalize $\rho$ by adding $f$ and elimi-}\\
      \>\>\> // \com{nating conjunct for one user attribute used in $f$}\\
\lnum\>\>\> $\rho' = \langle$\=$\uae(\rho)[\uattr(f)\mapsto\top], \rae(\rho),$\\
\lnum\>\>\>\> $\ops(\rho), \con(\rho)\union\set{f}\rangle$\\
\lnum\>\>\> \ifstmt\ $\mean{\rho'} \subseteq \up_0$\\
\lnum\>\>\>~~~~\= $\cc'.add(f)$\\
\lnum\>\>\>\>     $\gen.add(\rho')$\\
\lnum\>\>\> \elsestmt\\
 \>~~~~\= // \com{try to generalize $\rho$ by adding $f$ and elimi-}\\
     \>\>     // \com{nating conjunct for one resource attribute used in $f$.}\\
\lnum\>\>\>\> $\rho' = \langle$\=$\uae(\rho), \rae(\rho)[\rattr(f)\mapsto\top],$\\
\lnum\>\>\>\>\> $\ops(\rho), \con(\rho)\union\set{f}\rangle$\\
\lnum\>\>\>\> \ifstmt\ $\mean{\rho'} \subseteq \up_0$\\
\lnum\>\>\>\>~~~~\= $\cc'.add(f)$\\
\lnum\>\>\>\>\>     $\gen.add(\rho')$\\
\lnum\>\>\>\> \eifstmt\\
\lnum\>\>\> \eifstmt\\
\lnum\>\> \eifstmt\\
\lnum\> \eforloop\\
\lnum\> \forloop\ $i$ = 1 \forloopto\ $\cc'$.length\\
\lnum\>\> // \com{try to further generalize $\gen[i]$}\\
\lnum\>\> $\rho'' = \generalizeruleL($\=$\gen[i], \cc'[i\!+\!1\,..], \uncovup,$\\
\lnum\>\>\> $\Rho)$\\
\lnum\>\>     \ifstmt\ $\QrulL(\rho'', \uncovup) > \QrulL($\=$\rhobest, \uncovup)$\\
\lnum\>\>~~~~\= $\rhobest = \rho''$\\
\lnum\>\> \eifstmt\\
\lnum\> \eforloop\\
\lnum\> \return\ $\rhobest$
\end{tabbing}
\caption{Generalize rule $\rho$ by adding some formulas from $\cc$ to its
  constraint and eliminating conjuncts for attributes used in those
  formulas.  $f[x \mapsto y]$ denotes a copy of function $f$ modified so
  that $f(x)=y$.  $a[i..]$ denotes the suffix of array $a$ starting at
  index $i$. }
\label{fig:generalizerule}
\end{figure}


\begin{figure}[tbp]
\resetlnum
\begin{tabbing}
~~~~~\= \hypertarget{mergerules}{\function\ \mergerules(\Rho)}\\
\lnum\> // \com{Remove redundant rules}\\
\lnum\> $\rdtrules = \setc{\rho\in\Rho}{\exists\, \rho' \in \Rho\setminus\set{\rho}.\; \mean{\rho}\subseteq\mean{\rho'}}$\\
\lnum\> $\Rho.\removeall(\rdtrules)$\\
\lnum\> // \com{Merge rules}\\
\lnum\> $\wkset = \{(\rho_1,\rho_2) \;|\;$\=$\rho_1 \in \Rho \land \rho_2 \in \Rho$\\
\>\>                               ${} \land \rho_1\ne \rho_2 \land \con(\rho_1)=\con(\rho_2)\}$\\
\lnum\>\whileloop\ not(\wkset.empty())\\
\>~~~~\=      // \com{Remove an arbitrary element of the workset}\\
\lnum\>\>     $(\rho_1,\rho_2) = \wkset.{\rm remove}()$\\
\lnum\>\>     $\rhomerge = \langle$\=$\uae(\rho_1)\union\uae(\rho_2),
                         \rae(\rho_1)\union\rae(\rho_2),$\\
\>\>\>              $\ops(\rho_1)\union\ops(\rho_2),
                         \con(\rho_1)\rangle$\\
\lnum\>\>     \ifstmt\ $\mean{\rhomerge} \subseteq \up_0$\\
\>\>~~~~\=       // \com{The merged rule is valid.  Add it to $\Rho$,}\\
\>\>\>           // \com{and remove rules that became redundant.}\\
\lnum\>\>\>         $\rdtrules = \setc{\rho\in\Rho}{\mean{\rho}\subseteq\mean{\rhomerge}}$\\
\lnum\>\>\>        $\Rho.\removeall(\rdtrules)$\\
\lnum\>\>\>      $\wkset.\removeall(\{$\=$(\rho_1,\rho_2) \in \wkset \;|\; $\\
\>\>\>    $~~~~~~~~~~~~~~~~~{} \rho_1 \in \rdtrules \vee \rho_2 \in \rdtrules\})$\\
\lnum\>\>\>      $\wkset.\addall(\{$\=$(\rhomerge,\rho) \;|\; \rho \in \Rho$\\
\>\>\>\>    $~~~~~{} \land \con(\rho)=\con(\rhomerge)\})$\\
\lnum\>\>\>      $\Rho.\add(\rhomerge)$\\
\lnum\>\> \eifstmt\\
\lnum\>\ewhileloop\\
\lnum\>\return\ $\true$ if any rules were merged\\
\end{tabbing}
\caption{Merge pairs of rules in $\Rho$, when possible, to reduce the WSC of
  $\Rho$.  $(a,b)$ denotes an unordered pair with components $a$ and $b$.
  The union $e=e_1 \union e_2$ of attribute expressions $e_1$ and $e_2$
  over the same set $A$ of attributes is defined by: for all attributes $a$
  in $A$, if $e_1(a)=\top$ or $e_2(a)=\top$ then $e(a)=\top$ otherwise
  $e(a)=e_1(a)\union e_2(a)$.}
\label{fig:mergerules}
\end{figure}

The function $\mergerulesL(\Rho)$ in Figure \ref{fig:mergerules} attempts to
reduce the WSC of $\Rho$ by removing redundant rules and merging pairs of
rules. A rule $\rho$ in $\Rho$ is {\em redundant} if $\Rho$ contains
another rule $\rho'$ such that $\mean{\rho}\subseteq\mean{\rho'}$.
Informally, rules $\rho_1$ and $\rho_2$ are merged by taking, for each
attribute, the union of the conjuncts in $\rho_1$ and $\rho_2$ for that
attribute.  If the resulting rule $\rhomerge$ is valid, $\rhomerge$ is
added to $\Rho$, and $\rho_1$ and $\rho_2$ and any other rules that are now
redundant are removed from $\Rho$.  $\mergerulesL(\Rho)$ updates its argument
$\Rho$ in place, and it returns a Boolean indicating whether any rules were
merged.

\shortonly{The function \hypertarget{simplifyrules}{$\simplifyrules(\Rho)$}
  attempts to simplify all of the rules in $\Rho$.  It updates its argument
  $\Rho$ in place, replacing rules in $\Rho$ with simplified versions when
  simplification succeeds.  It returns a Boolean indicating whether any
  rules were simplified.  It attempts to simplify each rule in the
  following ways.  (1) It eliminates sets that are supersets of other sets
  in conjuncts for multi-valued user attributes.  The $\supseteq$-based
  semantics for such conjuncts implies that this does not change the
  meaning of the conjunct.  For example, a conjunct $\set{\set{a},
    \set{a,b}}$ is simplified to $\set{\set{a}}$.  (2) It eliminates
  elements from sets in conjuncts for multi-valued user attributes when
  this preserves validity of the rule; note that this might increase but
  cannot decrease the meaning of a rule.  For example, if every user whose
  specialties include $a$ also have specialty $b$, and a rule contains the
  conjunct $\set{\set{a,b}}$ for the specialties attribute, then $b$ will
  be eliminated from that conjunct.  (3) It eliminates conjuncts from a
  rule when this preserves validity of the rule.  Since removing one
  conjunct might prevent removal of another conjunct, it searches for the
  set of conjuncts to remove that maximizes the quality of the resulting
  rule, while preserving validity.  The user can specify a set of {\em
    unremovable attributes}, i.e., attributes for which $\simplifyrulesL$
  should not try to eliminate the conjunct, because eliminating it would
  increase the risk of generating an overly general policy, i.e., a policy
  that might grant inappropriate permissions when new users or new
  resources (hence new permissions) are added to the system.  Our
  experience suggests that appropriate unremovable attributes can be
  identified based on the obvious importance of some attributes and by
  examination of the policy generated without specification of unremovable
  attributes.  (4) It eliminates atomic constraints from a rule when this
  preserves validity of the rule.  It searches for the set of atomic
  constraints to remove that maximizes the quality of the resulting rule,
  while preserving validity.  (5) It eliminates overlapping values between
  rules.  Specifically, a value $v$ in the conjunct for a user attribute
  $a$ in a rule $\rho$ is removed if there is another rule $\rho'$ in the
  policy such that ({\it i}) $\attrib(\uae(\rho')) \subseteq
  \attrib(\uae(\rho))$ and $\attrib(\rae(\rho')) \subseteq
  \attrib(\rae(\rho))$, ({\it ii}) the conjunct of $\uae(\rho')$ for $a$
  contains $v$, ({\it iii}) each conjunct of $\uae(\rho')$ or $\rae(\rho')$
  other than the conjunct for $a$ is either $\top$ or a superset of the
  corresponding conjunct of $\rho$, and ({\it iv}) $\con(\rho') \subseteq
  \con(\rho)$.  The condition for removal of a value in the conjunct for a
  resource attribute is analogous.  If a conjunct of $\uae(\rho)$ or
  $\rae(\rho)$ becomes empty, $\rho$ is removed from the policy.  For
  example, if a policy contains the rules $\langle \dept \in \set{d_1,d_2}
  \land \position=p_1, \type=t_1, {\rm read}, \dept=\dept \rangle$ and
  $\langle \dept \in \set{d_1} \land \position=p_1, \type\in \set{t_1,t_2},
  {\rm read}, \dept=\dept \rangle$, then $d_1$ is eliminated from the
  former rule.  (6) It eliminates overlapping operations between rules.
  The details are similar to those for elimination of overlapping values
  between rules.  For example, if a policy contains the rules $\langle
  \dept=d_1, \type=t_1, {\rm read}, \dept=\dept \rangle$ and $\langle
  \dept=d_1 \land \position=p_1, \type=t_1, \set{{\rm read}, {\rm write}},
  \dept=\dept \rangle$, then {\rm read} is eliminated from the latter rule.
} %

\myparagraph{Asymptotic Running Time}



The algorithm's overall running time is worst-case cubic in $|\up_0|$.  A
detailed analysis of the\fullonly{ worst-case} asymptotic running time
appears in Section \ref{sec:aymptotic-time} in the Supplemental
Material.\fullonly{ Briefly, in the worst case, the algorithm generates one
  rule per user-permission tuple, and $\mergerules$ is worst-case cubic in
  the number of rules.}  In the experiments with sample policies and
synthetic policies described in Section \ref{sec:evaluation}, the observed
running time is roughly quadratic and roughly linear, respectively, in
$|\up_0|$.


\myparagraph{Attribute Selection}

Attribute data may contain attributes irrelevant to access control.  This
potentially hurts the effectiveness and performance of policy mining
algorithms \cite{frank09probabilistic,ni09automating}.
Therefore, before applying our algorithm to a dataset that might contain
irrelevant attributes, it is advisable to use the method in
\cite{frank09probabilistic} or \cite{molloy10noisy} to determine the
relevance of each attribute to the user-permission assignment and then
eliminate attributes with low relevance.

\myparagraph{Processing Order}

The order in which tuples and rules are processed can affect the mined
policy.  The order in which our algorithm processes tuples and rules is
described in Section \ref{sec:processing-order} in the Supplemental
Material.

\myparagraph{Optimizations}

Our implementation incorporates a few optimizations not reflected in the
pseudocode but described in Section \ref{sec:optimizations} in the
Supplemental Material.  The most novel optimization is that rules are
merged (by calling $\mergerulesL$) periodically, not only after all of
$\up_0$ has been covered.  This is beneficial because merging sometimes has
the side-effect of generalization, which causes more user-permission tuples
to be covered without explicitly considering them as seeds.

\subsection{Noise Detection}
\label{sec:algorithm:noise}

In practice, the given user-permission relation often contains noise,
consisting of over-assignments and under-assignments. An {\em
  over-assignment} is when a permission is inappropriately granted to a
user.  An {\em under-assignment} is when a user lacks a permission that he
or she should be granted.  Noise incurs security risks and significant IT
support effort \cite{molloy10noisy}.  This section describes extensions of
our algorithm to handle noise.  The extended algorithm detects and reports
suspected noise and generates an ABAC policy that is consistent with its
notion of the correct user-permission relation (i.e., with the suspected
noise removed).  The user should examine the suspected noise and decide
which parts of it are actual noise (i.e., errors in the user-permission
relation).  If all of it is actual noise, then the policy already generated
is the desired one; otherwise, the user should remove the parts that are
actual noise from the user-permission relation to obtain a correct
user-permission relation and then run the algorithm without the noise
detection extension on it to generate the desired ABAC policy.

Over-assignments are often the result of incomplete revocation of old
permissions when users change job functions \cite{molloy10noisy}.
Therefore, over-assignments usually cannot be captured concisely using
rules with attribute expressions that refer to the current attribute
information, so a candidate rule constructed from a user-permission tuple
that is an over-assignment is less likely to be generalized and merged with
other rules, and that candidate rule will end up as a low-quality rule in
the generated policy.  So, to detect over-assignments, we introduce a rule
quality threshold $\tau$.  The rule quality metric used here is the first
component of the metric used in the loop in Figure \ref{fig:alg} that
constructs $\Rho'$; thus, $\tau$ is a threshold on the value of
$\QrulL(\rho, \uncovup)$, and the rules with quality less than or equal to
$\tau$ form a suffix of the sequence of rules added to $\Rho'$.  
The extended algorithm reports as suspected over-assignments the
user-permission tuples covered in $\Rho'$ only by rules with quality less
than or equal to $\tau$, and then it removes rules with quality less than
or equal to $\tau$ from $\Rho'$.  Adjustment of $\tau$ is guided by the
user.  For example, the user might guess a percentage of over-assignments
(e.g., 3\%) based on experience, and let the system adjust $\tau$ until the
number of reported over-assignments is that percentage of $|\up_0|$.  

To detect under-assignments, we look for rules that are almost valid, i.e.,
rules that would be valid if a relatively small number of tuples were added
to $\up_0$.  A parameter $\alpha$ quantifies the notion of ``relatively
small''.  A rule is {\em $\alpha$ almost valid} if the fraction of invalid
user-permission tuples in $\mean{\rho}$ is at most $\alpha$, i.e.,
$|\mean{\rho}\setminus\up_0| \div |\mean{\rho}| \leq \alpha$.  In places
where the policy mining algorithm checks whether a rule is valid,
if the rule is $\alpha$ almost valid, the algorithm treats it as if it were
valid.  The extended algorithm reports $\UNION_{\rho \in \Rho'}
\mean{\rho}\setminus \up_0$ as the set of suspected under-assignments, and
(as usual) it returns $\Rho'$ as the generated policy.  Adjustment of
$\alpha$ is guided by the user, similarly as for the over-assignment
threshold $\tau$.


\section{Evaluation}
\label{sec:evaluation}

The general methodology used for evaluation is described in Section
\ref{sec:intro}.  We applied this methodology to sample policies and
synthetic policies.  Evaluation on policies (including attribute data) from
real organizations would be ideal, but we are not aware of any suitable and
publicly available policies from real organizations.  Therefore, we
developed sample policies that, although not based directly on specific
real-world case studies, are intended to be similar to policies that might
be found in the application domains for which they are named.  The sample
policies are relatively small and intended to resemble interesting core
parts of full-scale policies in those application domains.  Despite their
modest size, they are a significant test of the effectiveness of our
algorithm, because they express non-trivial policies and exercise all
features of our policy language, including use of set membership and
superset relations in attribute expressions and constraints.  The synthetic
policies are used primarily to assess the behavior of the algorithm as a
function of parameters controlling specific structural characteristics of
the policies.

We implemented our policy mining algorithm in Java and ran experiments on a
laptop with a 2.5 GHz Intel Core i5 CPU.  All of the code and data is
available at \url{http://www.cs.sunysb.edu/~stoller/}.  In our experiments,
the weights $w_i$ in the definition of WSC equal~1.

\subsection{Evaluation on Sample Policies}
\label{sec:eval:sample-policies}

We developed four sample policies, each consisting of rules and a manually
written attribute dataset containing a small number of instances of each
type of user and resource.  We also generated synthetic attribute datasets
for each sample policy.  The sample policies are described very briefly in
this section.  Details of the sample policies, including all policy rules,
some illustrative manually written attribute data, and a more detailed
description of the synthetic attribute data generation algorithm appear in
Section \ref{sec:sample-policy-details} in the Supplemental Material.

Figure \ref{fig:sample-policies-size} provides information about their
size.  Although the sample policies are relatively small when measured by a
coarse metric such as number of rules, they are complex, because each rule
has a lot of structure.  For example, the number of well-formed rules built
using the attributes and constants in each policy and that satisfy the
strictest syntactic size limits satisfied by rules in the sample policies
(at most one conjunct in each UAE, at most two conjuncts in each RAE, at
most two atomic constraints in each constraint, at most one atomic value in
each UAE conjunct, at most two atomic values in each RAE conjunct, etc.) is
more than $10^{12}$ for the sample policies with manually written attribute
data and is much higher for the sample policies with synthetic attribute
data and the synthetic policies.

In summary, our algorithm is very effective for all three sample policies:
there are only small differences between the original and mined policies if
no attributes are declared unremovable, and the original and mined policies
are identical if the resource-type attribute is declared unremovable.





\begin{figure*}
  \begin{center}
\begin{tabular}[tb]{|l|l|l|l|l|l|l|l|l|l|l|l|}
\hline
Policy & $|\Rho|$ & $|\Au|$ & $|\Ar|$ & $|\Op|$ & Type & $N$ & $|U|$ & $|R|$ & $|\vals|$ & $|\up|$ & $\avgUPrule$ \\ 
\hline
university & 10 & 6 & 5 & 9 & man & 2 & 22 & 34 & 76 & 168 & 19 \\ 
\hline
 &  &  &  &  & syn & 10 & 479 & 997 & 1651 & 8374 & 837 \\ 
\hline
 &  &  &  &  & syn & 20 & 920 & 1918 & 3166 & 24077 & 2408 \\ 
\hline
health care & 9 & 6 & 7 & 3 & man & 2 & 21 & 16 & 55 & 51 & 6.7 \\ 
\hline
 &  &  &  &  & syn & 10 & 200 & 720 & 1386 & 1532 & 195 \\ 
\hline
 &  &  &  &  & syn & 20 & 400 & 1440 & 2758 & 3098 & 393 \\ 
\hline
project mgmt & 11 & 8 & 6 & 7 & man & 2 & 19 & 40 & 77 & 189 & 19 \\ 
\hline
 &  &  &  &  & syn & 10 & 100 & 200 & 543 & 960 & 96 \\ 
\hline
 &  &  &  &  & syn & 20 & 200 & 400 & 1064 & 1920 & 193 \\ 
\hline
\end{tabular}
\video{viedo & ADD DATA HERE}
  \end{center}
    \caption{Sizes of the sample policies.  ``Type'' indicates whether the
      attribute data in the policy is manually written (``man'') or
      synthetic (``syn'').  $N$ is the number of departments for the
      university and project management sample policies, and the number of
      wards for the health care sample policy.  $\avgUPrule$ is the average
      number of user-permission tuples that satisfy each rule. 
      An empty cell indicates the same value as the cell above it.
    }
    \label{fig:sample-policies-size}
\end{figure*}

\myparagraph{University Sample Policy}

Our university sample policy controls access by students,
instructors, teaching assistants, registrar officers, department chairs,
and admissions officers to applications (for admission), gradebooks,
transcripts, and course schedules.  If no attributes are declared
unremovable, the generated policy is the same as the original ABAC policy
except that the RAE conjunct ``type=transcript'' is replaced with the
constraint ``department=department'' in one rule.
If resource type is declared unremovable, the generated policy is identical
to the original ABAC policy.

\myparagraph{Health Care Sample Policy}

Our health care sample policy controls access by nurses, doctors, patients,
and agents (e.g., a patient's spouse) to electronic health records (HRs)
and HR items (i.e., entries in health records).  If no attributes are
declared unremovable, the generated policy is the same as the original ABAC
policy except that the RAE conjunct ``type=HRitem'' is eliminated from four
rules; that conjunct is unnecessary, because those rules also contain a
conjunct for the ``topic'' attribute, and the ``topic'' attribute is used
only for resources with type=HRitem.
If resource type is declared unremovable, the generated policy is
identical to the original ABAC policy.

\myparagraph{Project Management Sample Policy}

Our project management sample policy controls access by department
managers, project leaders, employees, contractors, auditors, accountants,
and planners to budgets, schedules, and tasks associated with projects.  If
no attributes are declared unremovable, the generated policy is the same as
the original ABAC policy except that the RAE conjunct ``type=task'' is
eliminated from three rules; the explanation is similar to the above
explanation for the health care sample policy.
If resource type is declared  unremovable, the generated policy is
identical to the original ABAC policy.

\video{\myparagraph{Online Video Sample Policy}

Our online video sample policy controls access to videos by users of an
online video service.  The generated policy is identical to the
``original'' ABAC policy presented in Section \ref{sec:sample-policy-details}
in the Supplemental Material.  That is not surprising, because the
``original'' policy was actually produced by applying our algorithm to a
simple-minded ACL-like policy, which contains, for each combination of
video type and rating, one rule specifying the membership categories and
age groups permitted to view those videos.  The generated policy is much
more concise than the ACL-like policy (WSC=20 {\it vs.} WSC=39) and
generalizes better.  For example, if a new rating, such as PG-13, is
introduced, the minimized policy automatically grants adults permission to
view movies with the new rating; the ACL-like policy does not.
}

\myparagraph{Running Time on Synthetic Attribute Data}

\begin{figure}[htb]
  \centering
  \includegraphics[width=85mm]{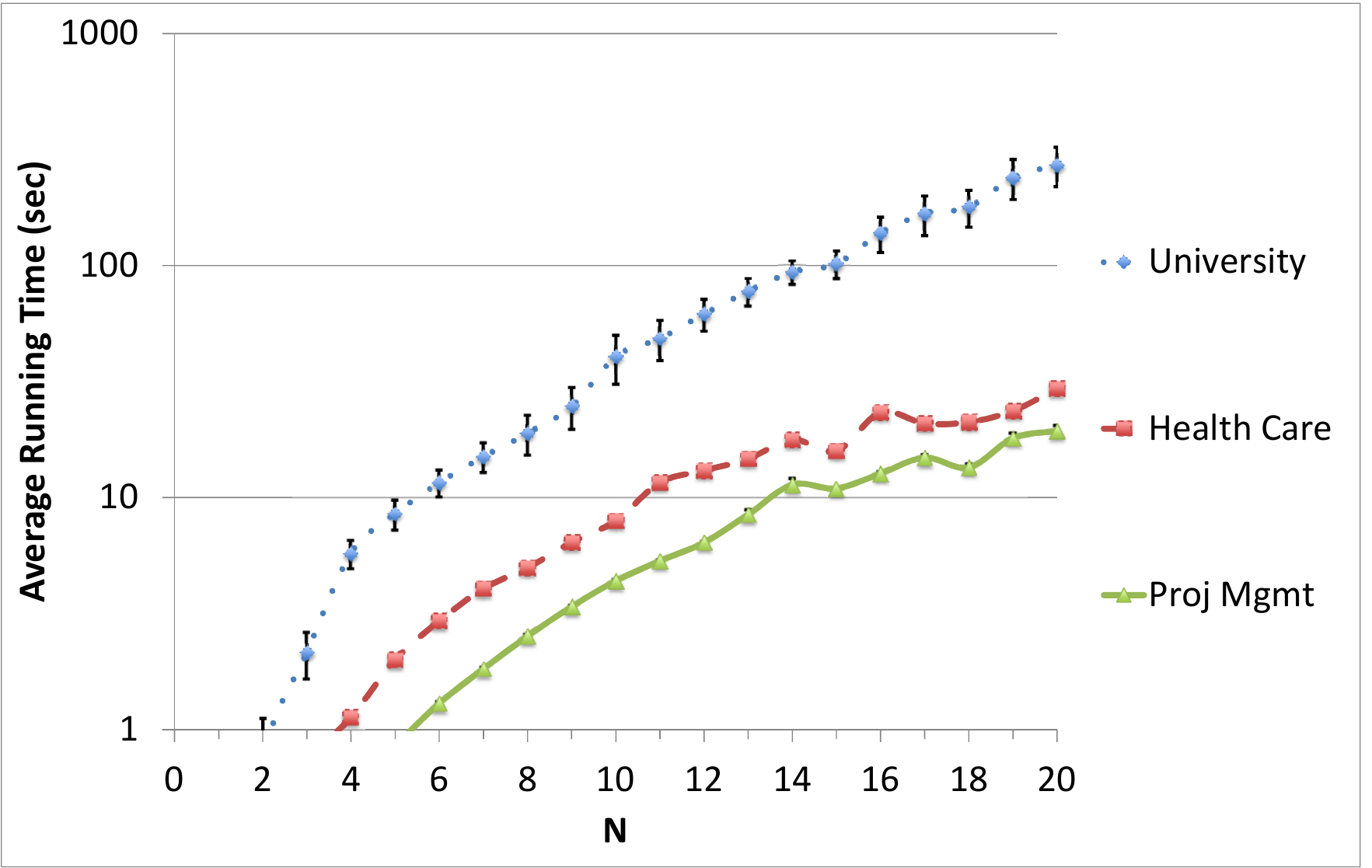}
  \caption{Running time (log scale) of the algorithm on synthetic attribute
    datasets for sample policies.  The horizontal axis is $\Ndept$ for 
    university and project management sample policies and $\Nward$ for
    health care sample policy. }
  \label{fig:sample-policies-time}
\end{figure}

We generated a series of pseudorandom synthetic attribute datasets for
the\video{ first three} sample policies, parameterized by a number $N$,
which is the number of departments for the university and project
management sample policies, and the number of wards for the health care
sample policy.  The generated attribute data for users and resources
associated with each department or ward are similar to but more numerous
than the attribute data in the manually written datasets.  Figure
\ref{fig:sample-policies-size} contains information about the sizes of the
policies with synthetic attribute data, for selected values of $N$.
Policies for the largest shown value of $N$ are generated as described in
Section \ref{sec:sample-policy-details} in the Supplemental Material;
policies for smaller values of $N$ are prefixes of them.  Each row contains
the average over 20 synthetic policies with the specified $N$.
For all sizes of synthetic attribute data, the mined policies are the same
as with the manually generated attribute data.  This reflects that larger
attribute datasets are not necessarily harder to mine from, if they
represent more instances of the same rules; the complexity is primarily in
the structure of the rules.  Figure \ref{fig:sample-policies-time} shows
the algorithm's running time as a function of $N$.  Each data point is an
average of the running times on 20 policies with synthetic attribute data.
Error bars (too small to see in most cases) show 95\% confidence intervals
using Student's t-distribution.  The running time is a roughly quadratic
function of $N$ for all three sample policies, with different constant
factors.  Different constant factors are expected, because policies are
very complex structures, and $N$ captures only one aspect of the size and
difficulty of the policy mining problem instance.  For example, the
constant factors are larger for the university sample policy mainly because
it has larger $|\up|$, as a function of $N$, than the other sample
policies.  For example, Figure \ref{fig:sample-policies-size} shows that
$|\up|$ for the university sample policy with $N=10$ is larger than $|\up|$
for the other sample policies with $N=20$.




\myparagraph{Benefit of Periodic Rule Merging Optimization}


It is not obvious {\it a priori} whether the savings from periodic merging
of rules outweighs the cost.  In fact, the net benefit grows with policy
size.  For example, for the university policy with synthetic attribute
data, this optimization provides a speedup of (67 sec)/(40 sec) = 1.7 for
$\Ndept=10$ and a speedup of (1012 sec)/(102 sec) = 9.9 for $\Ndept=15$.




\subsection{Evaluation on Synthetic Policies}
\label{sec:eval:synthetic}

We also evaluated our algorithm on synthetic ABAC policies.  On the
positive side, synthetic policies can be generated in all sizes and with
varying structural characteristics.  On the other hand, even though our
synthesis algorithm is designed to generate policies with some realistic
characteristics, the effectiveness and performance of our algorithm on
synthetic policies might not be representative of its effectiveness and
performance on real policies.
For experiments with synthetic policies, we compare the syntactic
similarity and $\wsc$ of the synthetic ABAC policy and the mined ABAC
policy.  Syntactic similarity of policies measures the syntactic similarity
of rules in the policies.  It ranges from 0 (completely different) to 1
(identical).  The detailed definition of syntactic similarity is in Section
\ref{sec:syntactic-similarity} in the Supplemental Material.  We do not
expect high syntactic similarity between the synthetic and mined ABAC
policies, because synthetic policies tend to be unnecessarily complicated,
and mined policies tend to be more concise.  Thus, we consider the policy
mining algorithm to be effective if the mined ABAC policy $\Rho_{\rm
  mined}$ is simpler (i.e., has lower $\wsc$) than the original synthetic
ABAC policy $\Rho_{\rm syn}$.  We compare them using the {\em compression
  factor}, defined as $\wsc(\Rho_{\rm syn})/\wsc(\Rho_{\rm mined})$.  Thus,
a compression factor above 1 is good, and larger is better.

\myparagraph{Synthetic Policy Generation}

Our policy synthesis algorithm first generates the rules and then uses the
rules to guide generation of the attribute data; this allows control of the
number of granted permissions.  Our synthesis algorithm takes $\Nrule$, the
desired number of rules, $\Ncnj$, the minimum number of conjuncts in each
attribute expression, and $\Ncns$, the minimum number of constraints in
each rule, as inputs. The numbers of users and resources are not specified
directly but are proportional to the number of rules, since our algorithm
generates new users and resources to satisfy each generated rule, as
sketched below. Rule generation is based on several statistical
distributions, which are either based loosely on our sample policies or
assumed to have a simple functional form (e.g., uniform distribution or
Zipf distribution).
For example, the distribution of the number of conjuncts in each attribute
expression is based loosely on our sample policies and ranges from $\Ncnj$
to $\Ncnj+3$, the distribution of the number of atomic constraints in each
constraint is based loosely on our sample policies and ranges from $\Ncns$
to $\Ncns+2$, and the distribution of attributes in attribute expressions
is assumed to be uniform (i.e., each attribute is equally likely to be
selected for use in each conjunct). 

The numbers of user attributes and resource attributes are fixed at $N_{\rm
  attr}=8$ (this is the maximum number of attributes relevant to access
control for the datasets presented in \cite{molloy11adversaries}).  Our
synthesis algorithm adopts a simple type system, with 7 types, and with at
least one user attribute and one resource attribute of each type.  For each
type $t$, the cardinality $c(t)$ is selected from a uniform distribution on
the interval $[2, 10\Nrule+2]$, the target ratio between the frequencies of
the most and least frequent values of type $t$ is chosen to be 1, 10, or
100 with probability 0.2 0.7, and 0.1, respectively, and a skew $s(t)$ is
computed so that the Zipf distribution with cardinality $c(t)$ and skew
$s(t)$ has that frequency ratio.  When assigning a value to an attribute of
type $t$, the value is selected from the Zipf distribution with cardinality
$c(t)$ and skew $s(t)$.  Types are also used when generating constraints:
constraints relate attributes with the same type.


For each rule $\rho$, our algorithm ensures that there are at least $N_{\rm
  urp}=16$ user-resource pairs $\tuple{u,r}$ such that $\tuple{u, r, o}
\models \rho$ for some operation $o$.  The algorithm first checks how many
pairs of an existing user and an existing resource (which were generated
for previous rules) satisfy $\rho$ or can be made to satisfy $\rho$ by
appropriate choice of values for attributes with unknown values (i.e.,
$\bot$).  If the count is less than $N_{\rm urp}$, the algorithm generates
additional users and resources that together satisfy $\rho$.  With the
resulting modest number of users and resources, some conjuncts in the UAE
and RAE are likely to be unnecessary (i.e., eliminating them does not grant
additional permissions to any existing user).  In a real policy with
sufficiently large numbers of users and resources, all conjuncts are likely
be to necessary.  To emulate this situation with a modest number of users,
for each rule $\rho$, for each conjunct $e_u(a_u)$ in the UAE $e_u$ in
$\rho$, the algorithm generates a user $u'$ by copying an existing user $u$
that (together with some resource) satisfies $\rho$ and then changing
$\uad(u', a_u)$ to some value not in $e_u(a_u)$.
Similarly, the algorithm adds resources to increase the chance that
conjuncts in resource attribute expressions are necessary, and it adds
users and resources to increase the chance that constraints are necessary.
The algorithm initially assigns values only to the attributes needed to
ensure that a user or resource satisfies the rule under consideration.  To
make the attribute data more realistic, a final step of the algorithm
assigns values to additional attributes until the fraction of attribute
values equal to $\bot$ reaches a target fraction $\botfrac = 0.1$.




\myparagraph{Results for Varying Number of Conjuncts}

To explore the effect of varying the number of conjuncts, we generated
synthetic policies with $\Nrule$ ranging from 10 to 50 in steps of 20, with
$\Ncnj$ ranging from 4 to 0, and with $\Ncns=0$.  For each value of
$\Nrule$, synthetic policies with smaller $\Ncnj$ are obtained by removing
conjuncts from synthetic policies with larger $\Ncnj$.  For each
combination of parameter values (in these experiments and the experiments
with varying number of constraints and varying overlap between rules), we
generate 50 synthetic policies and average the results.  Some experimental
results appear in Figure \ref{fig:varying-conjuncts}.  For each value of
$\Nrule$, as the number of conjuncts decreases, $|\up|$ increases (because
the numbers of users and resources satisfying each rule increase), the
syntactic similarity increases (because as there are fewer conjuncts in
each rule in the synthetic policy, it is more likely that the remaining
conjuncts are important and will also appear in the mined policy), and the
compression factor decreases (because as the policies get more similar, the
compression factor must get closer to 1).  For example, for $\Nrule=50$, as
$\Ncnj$ decreases from 4 to 0, average $|\up|$ increases from 1975 to
11969, average syntactic similarity increases from 0.62 to 0.75, and
average compression factor decreases from 1.75 to 0.84.  The figure also
shows the density of the policies, where the density of a policy is defined
as $|\up|\div (|U|\times|P|)$, where the set of granted permissions is $P =
\UNION_{\tuple{u,r,o}\in\up} \set{\tuple{r,o}}$.  The average densities all
fall within the range of densities seen in the 9 real-world datasets shown
in \cite[Table 1]{molloy09evaluating}, namely, 0.003 to 0.19.
Density is a decreasing function of $\Nrule$, because $|\up|$, $|U|$, and
$|P|$ each grow roughly linearly as functions of $\Nrule$.
The standard deviations of some quantities are relatively large in some
cases, but, as the relatively small confidence intervals indicate, this is
due to the intrinsic variability of the synthetic policies generated by our
algorithm, not due to insufficient samples.



\begin{figure*}[tb]
\begin{center}
\begin{tabular}{|l|l|l|l|l|l|l|l|l|l|l|l|l|l|l|l|l|l|l|l|l|}
\hline
$\Nrule$ & $\Ncnj$ & \multicolumn{2}{c|}{$|U|$} & \multicolumn{2}{c|}{$|R|$} & \multicolumn{2}{c|}{$|\up|$} & \multicolumn{2}{c|}{$\avgUPrule$} & \multicolumn{2}{c|}{Density} & \multicolumn{3}{c|}{Synt. Sim.} & \multicolumn{3}{c|}{Compression} & \multicolumn{3}{c|}{Time} \\ 
\cline{3-21}
 &  & \multicolumn{1}{c|}{\meanh} & \stdv & \meanh & \stdv & \meanh & \stdv & \meanh & \stdv & \meanh & \stdv & \meanh & \stdv & CI & \meanh & \stdv & CI & \meanh & \stdv & CI \\ 
\hline
10 & 4 & 206 & 16 & 62 & 5.7 & 401 & 90 & 35 & 9.2 & .069 & .01 & .63 & 0.04 & 0.011 & 1.79 & .21 & .060 & 0.30 & 0.31 & 0.09 \\ 
\hline
 & 2 &  &  &  &  & 620 & 116 & 136 & 29 & .076 & .01 & .69 & 0.03 & 0.009 & 1.55 & .18 & .051 & 0.47 & 0.19 & 0.05 \\ 
\hline
 & 0 &  &  &  &  & 1025 & 222 & 298 & 62 & .081 & .01 & .78 & 0.05 & 0.014 & 1.12 & .16 & .045 & 1.02 & 0.42 & 0.12 \\ 
\hline
50 & 4 & 1008 & 38 & 318 & 15 & 1975 & 282 & 36 & 5.3 & .014 & .001 & .62 & 0.02 & 0.006 & 1.75 & .08 & .023 & 4.86 & 1.71 & 0.49 \\ 
\hline
 & 2 &  &  &  &  & 3314 & 526 & 144 & 20 & .015 & .001 & .68 & 0.02 & 0.006 & 1.52 & .08 & .023 & 11.78 & 3.41 & 0.97 \\ 
\hline
 & 0 &  &  &  &  & 11969 & 5192 & 438 & 136 & .025 & .007 & .75 & 0.03 & 0.009 & 0.84 & .18 & .051 & 58.88 & 18.7 & 5.31 \\ 
\hline
\end{tabular}
\end{center}
\caption{Experimental results for synthetic policies with varying $\Ncnj$.
  ``Synt. Sim.'' is syntactic similarity.  ``Compression'' is the
  compression factor.  \meanh\ is mean, \stdv\ is standard deviation, and
  CI is half-width of 95\% confidence interval using Student's
  $t$-distribution.  An empty cell indicates the same value as the cell
  above it.}
\label{fig:varying-conjuncts}
\end{figure*}


\begin{figure*}[tb]
\begin{center}
\begin{tabular}{|l|l|l|l|l|l|l|l|l|l|l|l|l|l|l|l|l|l|l|l|l|}
\hline
$\Nrule$ & $\Ncns$ & \multicolumn{2}{c|}{$|U|$} & \multicolumn{2}{c|}{$|R|$} & \multicolumn{2}{c|}{$|\up|$} & \multicolumn{2}{c|}{$\avgUPrule$} & \multicolumn{2}{c|}{Density} & \multicolumn{3}{c|}{Synt. Sim.} & \multicolumn{3}{c|}{Compression} & \multicolumn{3}{c|}{Time} \\ 
\cline{3-21}
 &  & \multicolumn{1}{c|}{\meanh} & \stdv & \meanh & \stdv & \meanh & \stdv & \meanh & \stdv & \meanh & \stdv & \meanh & \stdv & CI & \meanh & \stdv & CI & \meanh & \stdv & CI \\ 
\hline
10 & 2 & 172 & 19 & 54 & 6.5 & 529 & 162 & 49 & 15 & .084 & .020 & .72 & .04 & .011 & 1.50 & .14 & .040 & 0.28 & 0.17 & 0.05 \\ 
\hline
 & 1 &  &  &  &  & 679 & 174 & 105 & 32 & .093 & .018 & .72 & .05 & .014 & 1.43 & .18 & .051 & 0.36 & 0.17 & 0.05 \\ 
\hline
 & 0 &  &  &  &  & 917 & 325 & 172 & 57 & .110 & .034 & .70 & .05 & .014 & 1.30 & .20 & .057 & 0.49 & 0.20 & 0.06 \\ 
\hline
50 & 2 & 781 & 51 & 276 & 16 & 3560 & 596 & 61 & 9.9 & .020 & .003 & .67 & .02 & .006 & 1.29 & .14 & .040 & 12.72 & 3.95 & 1.12 \\ 
\hline
 & 1 &  &  &  &  & 5062 & 1186 & 137 & 28 & .024 & .004 & .66 & .02 & .006 & 1.15 & .14 & .040 & 16.78 & 5.09 & 1.45 \\ 
\hline
 & 0 &  &  &  &  & 8057 & 2033 & 241 & 56 & .031 & .006 & 0.64 & .02 & .006 & 0.96 & 0.13 & .037 & 23.38 & 6.78 & 1.93 \\ 
\hline
\end{tabular}
\end{center}
\caption{Experimental results for synthetic policies with varying
  $\Ncns$.}
\label{fig:varying-constraints}
\end{figure*}


\begin{figure*}[tb]
\begin{center}
\begin{tabular}{|l|l|l|l|l|l|l|l|l|l|l|l|l|l|l|l|l|l|l|l|l|}
\hline
$\Nrule$ & $\Pover$ & \multicolumn{2}{c|}{$|U|$} & \multicolumn{2}{c|}{$|R|$} & \multicolumn{2}{c|}{$|\up|$} & \multicolumn{2}{c|}{$\avgUPrule$} & \multicolumn{2}{c|}{Density} & \multicolumn{3}{c|}{Synt. Sim.} & \multicolumn{3}{c|}{Compression} & \multicolumn{3}{c|}{Time} \\ 
\cline{3-21}
 &  & \multicolumn{1}{c|}{\meanh} & \stdv & \meanh & \stdv & \meanh & \stdv & \meanh & \stdv & \meanh & \stdv & \meanh & \stdv & CI & \meanh & \stdv & CI & \meanh & \stdv & CI \\ 
\hline
50 & 0 & 693 & 37 & 247 & 15 & 5246 & 1445 & 92.3 & 30.0 & .029 & .006 & .74 & .02 & .006 & 1.16 & .11 & .03 & 12.39 & 6.34 & 1.80 \\ 
\hline
 & 0.5 & 655 & 53 & 216 & 16 & 7325 & 1838 & 333 & 69.6 & .039 & .008 & .73 & .03 & .009 & 1.18 & .21 & .06 & 11.64 & 4.95 & 1.41 \\ 
\hline
 & 1 & 664 & 58 & 216 & 16 & 7094 & 1473 & 563 & 109 & .038 & .006 & .71 & .03 & .009 & 1.23 & .29 & .08 & 11.34 & 4.08 & 1.16 \\ 
\hline
\end{tabular}
\end{center}
\caption{Experimental results for synthetic policies with varying
  $P_{over}$. }
\label{fig:varying-overlap}
\end{figure*}

\myparagraph{Results for Varying Number of Constraints}

To explore the effect of varying the number of constraints, we generated
synthetic policies with $\Nrule$ ranging from 10 to 50 in steps of 20, with
$\Ncns$ ranging from 2 to 0, and with $\Ncnj=0$.  For each value of
$\Nrule$, policies with smaller $\Ncns$ are obtained by removing
constraints from synthetic policies with larger $\Ncns$.  Some experimental
results appear in Figure \ref{fig:varying-constraints}.  For each value
of $\Nrule$, as the number of constraints decreases, $|\up|$ increases
(because the numbers of users and resources satisfying each rule increase),
syntactic similarity decreases (because our algorithm gives preference to
constraints over conjuncts, so when $\Ncns$ is small, the mined policy
tends to have more constraints and fewer conjuncts than the synthetic
policy), and the compression factor decreases (because the additional
constraints in the mined policy cause each rule in the mined policy to
cover fewer user-permission tuples on average, increasing the number of
rules and hence the $\wsc$).  For example, for $\Nrule=50$, as $\Ncns$
decreases from 2 to 0, average $|\up|$ increases from 3560 to 26472,
average syntactic similarity decreases slightly from 0.67 to 0.64, and
average compression factor decreases from 1.29 to 0.96.



\myparagraph{Results for Varying Overlap Between Rules}

We also explored the effect of varying overlap between rules, to test our
conjecture that policies with more overlap between rules are harder to
reconstruct through policy mining.  The {\it overlap} between rules
$\rho_1$ and $\rho_2$ is $\mean{\rho_1}\intersect\mean{\rho_2}$.  To
increase the average overlap between pairs of rules in a synthetic policy,
we extended the policy generation algorithm so that, after generating each
rule $\rho$, with probability $\Pover$ the algorithm generates another rule
$\rho'$ obtained from $\rho$ by randomly removing one conjunct from
$\uae(\rho)$ and adding one conjunct (generated in the usual way) to
$\rae(\rho)$; typically, $\rho$ and $\rho'$ have a significant amount of
overlap. We also add users and resources that together satisfy $\rho'$, so
that $\mean{\rho'}\not\subseteq\mean{\rho}$, otherwise $\rho'$ is
redundant. This construction is based on a pattern that occurs a few times
in our sample policies.  We generated synthetic policies with 30 rules,
using the extended algorithm described above. For each value of $\Nrule$,
we generated synthetic policies with $\Pover$ ranging from 0 to 1 in steps
of 0.25, and with $\Ncnj$ = 2 and $\Ncns$ = 0.  Some experimental results
appear in Figure \ref{fig:varying-overlap}.  For each value of $\Nrule$, as
$\Pover$ increases, the syntactic similarity decreases (because our
algorithm effectively removes overlap, i.e., produces policies with
relatively little overlap), and the compression factor increases (because
removal of more overlap makes the mined policy more concise).  For example,
for $\Nrule=50$, as $\Pover$ increases from 0 to 1, the syntactic
similarity decreases slightly from 0.74 to 0.71, and the compression factor
increases from 1.16 to 1.23.

\subsection{Generalization}
\label{sec:eval:generalization}

A potential concern with optimization-based policy mining algorithms is
that the mined policies might overfit the given data and hence not be
robust, i.e., not generalize well, in the sense that the policy requires
modifications to accommodate new users.  To evaluate how well policies
generated by our algorithm generalize, we applied the following
methodology, based on \cite{frank09probabilistic}.  The inputs to the
methodology are an ABAC policy mining algorithm, an ABAC policy $\pi$, and
a fraction $f$ (informally, $f$ is the fraction of the data used for
training); the output is a fraction $e$ called the {\em generalization
  error} of the policy mining algorithm on policy $\pi$ for fraction $f$.
Given a set $U'$ of users and a policy $\pi$, the associated resources for
$U'$ are the resources $r$ such that $\pi$ grants some user in $U'$ some
permission on $r$.  To compute the generalization error, repeat the
following procedure 10 times and average the results: randomly select a
subset $U'$ of the user set $U$ of $\pi$ with $|U'|/|U| = f$, randomly
select a subset $R'$ of the associated resources for $U'$ with $|R'|/|R| =
f$, generate an ACL policy $\piacl$ containing only the permissions for
users in $U'$ for resources in $R'$, apply the policy mining algorithm to
$\piacl$ with the attribute data to generate an ABAC policy $\pigen$,
compute the generalization error as the fraction of incorrectly assigned
permissions for users not in $U'$ and resources not in $R'$, i.e., as $|S
\ominus S'|/|S|$, where $S=\setc{\tuple{u,r,o}\in\mean{\pi}}{u\in
  U\setminus U' \land r\in R\setminus R'}$,
$S'=\setc{\tuple{u,r,o}\in\mean{\pi'}}{u\in U\setminus U' \land r\in
  R\setminus R'}$, and $\ominus$ is symmetric set difference.



We measured generalization error for $f$ from 0.1 to 0.5 in steps of 0.05
for the university (with $\Ndept=40$), health care (with $\Nward=40$), and
project management (with $\Ndept=40$) sample policies.  For the university
and health care sample policies, the generalization error is zero in all
these cases.  For the project management sample policy, the generalization
error is 0.11 at $f=0.1$, drops roughly linearly to zero at $f=0.35$, and
remains zero thereafter.  There are no other existing ABAC policy mining
algorithms, so a direct comparison of the generalization results from our
algorithm with generalization results from algorithms based on other
approaches, e.g., probabilistic models, is not currently possible.
Nevertheless, these results are promising and suggest that policies
generated by our algorithm generalize reasonably well.


\subsection{Noise}
\label{sec:eval:noise}

\myparagraph{Permission Noise}

To evaluate the effectiveness of our noise detection techniques in the
presence of permission noise, we started with an ABAC policy, generated an
ACL policy, added noise, and applied our policy mining algorithm to the
resulting policy.  To add a specified level $\nu$ of permission noise,
measured as a percentage of $|\up_0|$, we added $\nu |\up_0|/6$
under-assignments and $5 \nu |\up_0|/6$ over-assignments to the ACL policy
generated from the ABAC policy.  This ratio is based on the ratio of Type I
and Type II errors in \cite[Table 1]{molloy10noisy}.  
The over-assignments are user-permission tuples generated by selecting the
user, resource, and operation from categorical distributions with
approximately normally distributed probabilities (``approximately'' because
the normal distribution is truncated on the sides to have the appropriate
finite domain); we adopted this approach from \cite{molloy10noisy}.  The
under-assignments are removals of user-permission tuples generated in the
same way.  For each noise level, we ran our policy mining algorithm with
noise detection inside a loop that searched for the best values of $\alpha$
(considering values between $0.01$ and $0.09$ in steps of $.01$) and $\tau$
(considering 0.08, values between 0.1 and 0.9 in steps of 0.1, and between 1
and 10 in steps of 1), because we expect $\tau$ to depend on the noise
level, and we want to simulate an experienced administrator, so that the
results reflect the capabilities and limitations of the noise detection
technique rather than the administrator.  The best values of $\alpha$ and
$\tau$ are the ones that maximize the Jaccard similarity of the actual
(injected) noise and the reported noise.  ROC curves that illustrate the
trade-off between false positives and false negatives when tuning the
values of $\alpha$ and $\tau$ appear in Section \ref{sec:ROC} in the
Supplemental Material.





\begin{figure}[tb]
  \centering
  \includegraphics[width=85mm]{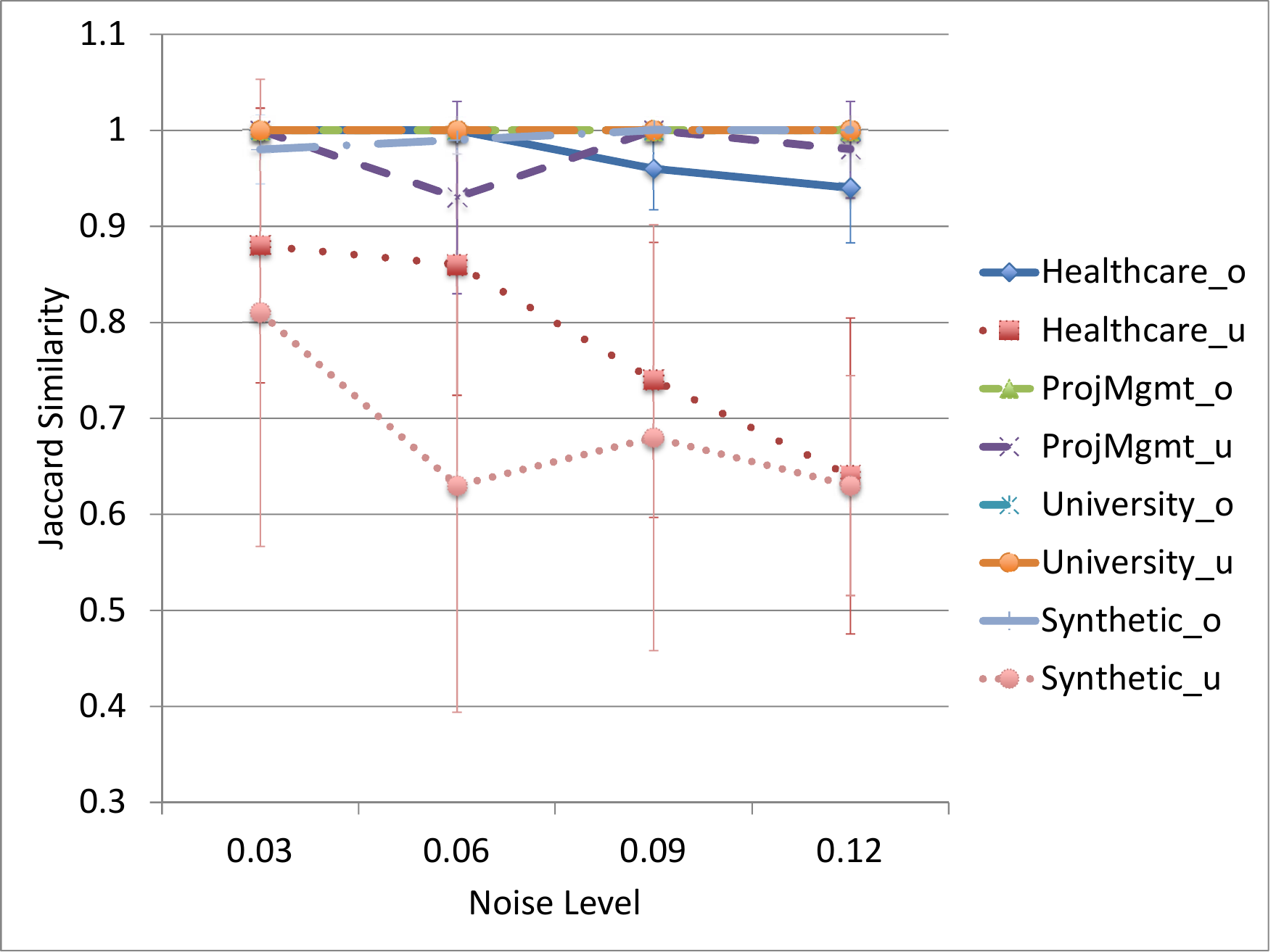}
  \caption{Jaccard similarity of actual and reported under-assignments, and
    Jaccard similarity of actual and reported over-assignments, as a
    function of permission noise level.  Curve names ending with $\_$o and
    $\_$u are for over-assignments and under-assignments, respectively.
    The curves for University$\_$u and Synthetic$\_$o are nearly the same
    and overlap each other. 
}
  \label{fig:jaccard-similarity-noise}
\end{figure}
\begin{figure}[tb]
  \centering
  \includegraphics[width=85mm]{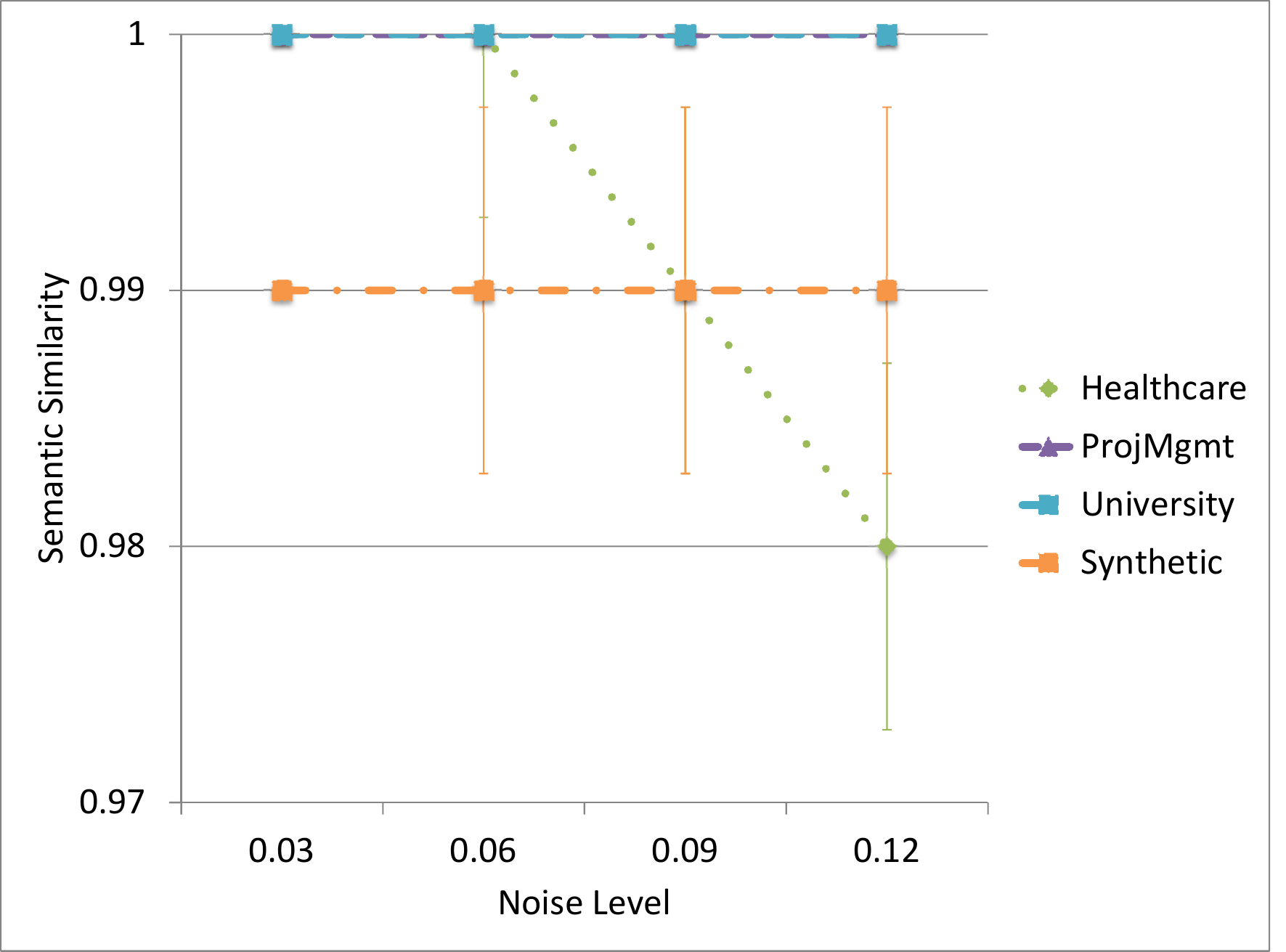}
  \caption{Semantic similarity of the original policy and the mined
    policy, as a function of permission noise level.}
  \label{fig:semantic-similarity-noise}
\end{figure}

We started with the university (with $\Ndept=4$), health care (with
$\Nward=6$), and project management (with $\Ndept=6$) sample policies with
synthetic attribute data (we also did some experiments with larger policy
instances and got similar results), and with synthetic policies with
$\Nrule=20$.  Figure \ref{fig:jaccard-similarity-noise} shows the Jaccard
similarity of the actual and reported over-assignments and the Jaccard
similarity of the actual and reported under-assignments.  Note that, for a
policy mining algorithm without noise detection (hence the reported noise
is the empty set), these Jaccard similarities would be 0.  Each data point
is an average over 10 policies, and error bars (too small to see in some
cases, and omitted when the standard deviation is 0) show 95\% confidence
intervals using Student's t-distribution.  Over-assignment detection is
accurate, with average Jaccard similarity always 0.94 or higher (in our
experiments).  Under-assignment detection is very good for university and
project management, with average Jaccard similarity always 0.93 or higher,
but less accurate for health care and synthetic policies, with average
Jaccard similarity always 0.63 or higher.  Intuitively, detecting
over-assignments is somewhat easier, because it is unlikely that there are
high-quality rules that cover the over-assignments, so we mostly get rules
that do not over-assign and hence the over-assignments get classified
correctly.  However, under-assignments are more likely to affect the
generated rules, leading to mis-classification of under-assignments.
As a function of noise level in the considered range, the Jaccard
similarities are flat in some cases and generally trend slightly downward
in other cases.  Figure \ref{fig:semantic-similarity-noise} shows the
semantic similarity of the original and mined policies.  Note that, for a
policy mining algorithm without noise detection, the semantic similarity
would equal $1-\nu$.  With our algorithm, the semantic similarity is always
significantly better than this.  The average semantic similarity is always
0.98 or higher, even for $\nu=0.12$.  The similarities are generally lower
for synthetic policies than sample policies, as expected, because synthetic
policies are not reconstructed as well even in the absence of noise.





\myparagraph{Permission Noise and Attribute Noise}

To evaluate the effectiveness of our noise detection techniques in the
presence of permission noise and attribute noise, we performed experiments
in which, for a given noise level $\nu$, we added $\nu |\up_0|/7$
under-assignments, $5 \nu |\up_0|/7$ over-assignments, and $\nu |\up_0|/7$
permission errors due to attribute errors to the ACL policy generated from
the ABAC policy (in other words, we add attribute errors until $\nu
|\up_0|/7$ user-permission tuples have been added or removed due to
attribute errors; this way, attribute errors are measured on the same scale
as under-assignments and over-assignments).  The attribute errors are
divided equally between missing values (i.e., replace a non-bottom value
with bottom) and incorrect values (i.e., replace a non-bottom value with
another non-bottom value).
Our current techniques do not attempt to distinguish permission noise from
attribute noise (this is a topic for future research); policy analysts are
responsible for determining whether a reported suspected error is due to an
incorrect permission, an incorrect or missing attribute value, or a false
alarm.  Since our techniques report only suspected under-assignments and
suspected over-assignments, when comparing actual noise to reported noise,
permission changes due to attribute noise (i.e., changes in the set of
user-permission tuples that satisfy the original policy rules) are included
in the actual noise.  We started with the same policies as above.  Graphs
of Jaccard similarity of actual and reported noise, and syntactic
similarity of original and mined policies, appear in Section
\ref{sec:attrib-noise-graphs} in the Supplemental Material.  The results
are similar to those without attribute noise, except with slightly lower
similarities for the same fraction of permission errors.  This shows that
our approach to noise detection remains appropriate in the presence of
combined attribute noise and permission noise.

\subsection{Comparison with Inductive Logic Programming}
\label{sec:eval:ILP}

We implemented a translation from ABAC policy mining to Inductive Logic
Programming (ILP) and applied Progol
\cite{muggleton00theory,muggleton01cprogol}, a well-known ILP system
developed by Stephen Muggleton, to translations of our sample policies and
synthetic policies.  Details of the translation appear in Section
\ref{sec:translation-to-ILP} in the Supplemental Material.  Progol mostly
succeeds in reconstructing the policies for university\video{, on-line
  video,} and project management, except it fails to learn rules with
conjuncts or operation sets containing multiple constants, instead
producing multiple rules.\fullonly{  For example, Progol generates four rules (one
for each combination of the constants) corresponding to the rule
$\langle${\tt true}, {\tt type} $\in$ \{{\tt schedule}, {\tt budget}\},
\{{\tt read}, {\tt write}\}, {\tt projectsLed} $\ni$ {\tt project}$\rangle$
in the project management case study.}  In addition, Progol fails to
reconstruct two rules in the health care sample policy.
Due to Progol's failure to learn rules with conjuncts or operation sets
containing multiple constants, we generated a new set of 20 synthetic
policies with at most 1 constant per conjunct and 1 operation per rule.
On these policies with $\Nrule=5$, our algorithm achieves a compression
factor of 1.92, compared to 1.67 for Progol.


Progol is much slower than our algorithm.  For the university (with
$\Ndept=10$), health care (with $\Nward=20$), and project management (with
$\Ndept=20$) sample policies, Progol is 302, 375, and 369 times slower than
our algorithm, respectively.  For synthetic policies with $\Nrule=5$,
Progol is 2.74 times slower than our algorithm; for synthetic policies with
$\Nrule=10$, we stopped Progol after several hours.


\section{Related Work}
\label{sec:related}


To the best of our knowledge, the algorithm in this paper is the first
policy mining algorithm for any ABAC framework.  Existing algorithms for
access control policy mining produce role-based policies; this includes
algorithms that use attribute data, e.g.,
\cite{molloy10mining,colantonio12decomposition,xu12algorithms}.  Algorithms
for mining meaningful RBAC policies from ACLs and user attribute data
\cite{molloy10mining,xu12algorithms} attempt to produce RBAC policies that
are small (i.e., have low $\wsc$) and contain roles that are meaningful in
the sense that the role's user membership is close to the meaning of some
user attribute expression.  User names (i.e., values of $\uid$) are used in
role membership definitions and hence are not used in attribute
expressions, so some sets of users cannot be characterized exactly by a
user attribute expression.  The resulting role-based policies are often
much larger than attribute-based policies, due to the lack of
parameterization; for example, they require separate roles for each
department in an organization, in cases where a single rule suffices in an
attribute-based policy.  Furthermore, algorithms for mining meaningful
roles does not consider resource attributes (or permission attributes),
constraints, or set relationships.

Xu and Stoller's work on mining parameterized RBAC (PRBAC) policies
\cite{xu13mining} is more closely related.  Their PRBAC framework supports
a simple form of ABAC, because users and permissions have attributes that
are implicit parameters of roles, the set of users assigned to a role is
specified by an expression over user attributes, and the set of permissions
granted to a role is specified by an expression over permission attributes.
Our work differs from theirs in both the policy framework and the
algorithm.  Regarding the policy framework, our ABAC framework supports a
richer form of ABAC than their PRBAC framework does.  Most importantly, our
framework supports multi-valued (also called ``set-valued'') attributes and
allows attributes to be compared using set membership, subset, and
equality; their PRBAC framework does not support multi-valued attributes,
and it allows attributes to be compared using only equality.  Multi-valued
attributes are very important in real policies.  Due to the lack of
multi-valued attributes, the sample policies in \cite{xu13mining} contain
artificial limitations, e.g., a faculty teaches only one course, and a
doctor is a member of only one medical team.  Our sample policies are
extensions of their case studies without these limitations: a faculty may
teach multiple courses, a doctor may be a member of multiple medical teams,
etc.
Our algorithm works in a different, and more efficient, way than theirs.
Our algorithm directly constructs rules to include in the output.
Their algorithm constructs a large set of candidate roles and then
determines which roles to include in the output, possibly discarding many
candidates (more than $90\%$ for their sample policies).

Ni {\it et al.} investigated the use of machine learning algorithms for
security policy mining \cite{ni09automating}.  Specifically, they use
supervised machine learning algorithms to learn classifiers that associate
permissions with roles, given as input the permissions, the roles,
attribute data for the permissions, and (as training data) the
role-permission assignment.  The resulting classifier---a support vector
machine (SVM)---can be used to automate assignment of new permissions to
roles.  They also consider a similar scenario in which a supervised machine
learning algorithm is used to learn classifiers that associate users with
roles, given as input the users, the roles, user attribute data, and the
user-role assignment.  The resulting classifiers are analogous to attribute
expressions, but there are many differences between their work and ours.
The largest difference is that their approach needs to be given the roles
and the role-permission or user-role assignment as training data; in
contrast, our algorithm does not require any part of the desired high-level
policy to be given as input.  Also, their work does not consider anything
analogous to constraints, but it could be extended to do so.
Exploring ABAC policy mining algorithms based on machine learning 
is a direction for future work.  

Lim {\it et al.} investigated the use of evolutionary algorithms to learn
and evolve security policies policies \cite{lim10evolving}.  They consider
several problems, including difficult problems related to risk-based
policies, but not general ABAC policy mining.  In the facet of their work
most similar to ABAC policy mining, they showed that genetic programming
can learn the access condition in the Bell-LaPadula multi-level security
model for mandatory access control.  The learned predicate was sometimes
syntactically more complex than, but logically equivalent to, the desired
predicate.


Association rule mining has been studied extensively.  Seminal work
includes Agrawal {\it et al.}'s algorithm for mining propositional rules
\cite{agrawal94fast}.  Association rule mining algorithms are not well
suited to ABAC policy mining, because they are designed to find rules that
are probabilistic in nature \cite{agrawal94fast} and are supported by
statistically strong evidence.  They are not designed to produce a set of
rules that are strictly satisfied, that completely cover the input data,
and are minimum-sized among such sets of rules.  Consequently, unlike our
algorithm, they do not give preference to smaller rules or rules with less
overlap (to reduce overall policy size).



Bauer {\it et al.} use association rule mining to detect policy errors
\cite{bauer08detecting}.  They apply propositional association rule mining
to access logs to learn rules expressing that a user who exercised certain
permissions is likely to exercise another permission.  A suspected
misconfiguration exists if a user who exercised the former permissions does
not have the latter permission.  
Bauer {\it et al.} do not consider attribute data or generate entire
policies.


Inductive logic programming (ILP) is a form of machine learning in which
concepts are learned from examples and expressed as logic programs.  ABAC
policies can be represented as logic programs, so ABAC policy mining can be
seen as a special case of ILP.  However, ILP systems are not ideally suited
to ABAC policy mining.  ILP is a more difficult problem, which involves
learning incompletely specified relations from a limited number of positive
and negative examples, exploiting background knowledge, etc.
ILP algorithms are correspondingly more complicated and less scalable, and
focus more on how much to generalize from the given examples than on
optimization of logic program size.  For example,  Progol ({\it cf.}
Section \ref{sec:eval:ILP})
uses a compression (rule size) metric to guide construction of each rule
but does not attempt to achieve good compression for the learned rules
collectively; in particular, it does not perform steps analogous to merging
rules, eliminating overlap between rules, and selecting the highest-quality
candidate rules for the final solution.  As the experiments in Section
\ref{sec:eval:ILP} demonstrate, Progol is slower and generally produces
policies with higher WSC, compared to our algorithm.


\section{Conclusion}
\label{sec:conclusion}

This paper presents an ABAC policy mining algorithm.  Experiments with
sample policies and synthetic policies demonstrate the algorithm's
effectiveness.  Directions for future work include supporting additional
ABAC policy language features
and exploring use of machine learning for ABAC policy mining.






\medskip
\myparagraph{Acknowledgments}

We thank the anonymous reviewers for the thorough and detailed reviews that
helped us improve this paper.

\bibliographystyle{IEEEtran}
\bibliography{IEEEabrv,../references}

\newcommand{\biographyvspace}{\vspace{-4ex}}

\biographyvspace
\begin{IEEEbiographynophoto}{Zhongyuan Xu}
  Zhongyuan Xu received the B.S. and M.S. degrees in Computer Science from
  Peking University in 2006 and 2009, respectively, and the Ph.D. degree in
  Computer Science from Stony Brook University in 2014.  He is currently a
  software developer at Facebook.
\end{IEEEbiographynophoto}

\biographyvspace
\begin{IEEEbiographynophoto}{Scott D. Stoller}
  Scott D. Stoller received the B.A. degree in Physics, summa cum laude,
  from Princeton University in 1990 and the Ph.D. degree in Computer
  Science from Cornell University in 1997.  He is currently a Professor 
  at Stony Brook University.  
\end{IEEEbiographynophoto}

\vfill

\clearpage

\section{Proof of NP-Hardness}
\label{sec:proof-NP-hard}

This section shows that the ABAC policy mining problem is NP-hard, by
reducing the Edge Role Mining Problem (Edge RMP) \cite{lu08optimal} to it.

An {\em RBAC policy} is a tuple $\pirbac=\tuple{U, P, R, \ua, \pa}$, where
$R$ is a set of roles, $\ua \subseteq U \times R$ is the user-role
assignment, and $\pa \subseteq R \times P$ is role-permission assignment.
The {\em number of edges} in an RBAC policy $\pirbac$ with this form is
$|\ua|+|\pa|$.  The user-permission assignment induced by an RBAC policy
with the above form is $\mean{\pirbac}=\ua \circ \pa$, where $\circ$
denotes relational composition.

The {\em Edge Role Mining Problem (Edge RMP)} is \cite{lu08optimal}: Given
an ACL policy $\tuple{U, P, \up}$, where $U$ is a set of users, $P$ is a
set of permissions, and $\up\subseteq U \times P$ is a user-permission
relation, find an RBAC policy $\pirbac=\tuple{U, P, R, \ua, \pa}$ such
that $\mean{\pirbac}=\up$ and $\pirbac$ has minimum number of edges among
RBAC policies satisfying this condition.  NP-hardness of Edge RMP follows
from Theorem 1 in \cite{molloy10mining}, since Edge RMP corresponds to
the Weighted Structural Complexity Optimization (WSCO) Problem with
$w_r=0$, $w_u=1$, $w_p=1$, $w_h=\infty$, and $w_d=\infty$.

Given an Edge RMP problem instance $\tuple{U, P, \up}$, consider the ABAC
policy mining problem instance with ACL policy $\pi_0=\tuple{U \union
  \set{u_0}, P \union \set{r_0}, \set{op_0}, \up_0}$, where $u_0$ is a new
user and $r_0$ is a new resource, $\up_0 =
\setc{\tuple{u,r,op_0}}{\tuple{u,r}\in\up}$, user attributes
$\Au=\set{\uid}$, resource attributes $\Ar=\set{\rid}$, user attribute
data $\uad$ defined by $\uad(u,\uid)=u$, resource attribute data $\rad$
defined by $\rad(r,\rid)=r$, and policy quality metric $\Qpol$ defined by
$\wscPolL$ with $w_1=1$, $w_2=1$, $w_3=0$, and $w_4=1$.  Without loss of
generality, we assume $U \intersect P = \emptyset$; this should always
hold, because in RBAC, users are identified by names that are atomic
values, and permissions are resource-operation pairs; if for some reason
this assumption doesn't hold, we can safely rename users or permissions to
satisfy this assumption, because RBAC semantics is insensitive to
equalities between users and permissions.

A solution to the given Edge-RMP problem instance can be constructed
trivially from a solution $\piabac$ to the above ABAC policy mining
instance by interpreting each rule as a role.  Note that rules in $\piabac$
do not contain any constraints, because $\uid$ and $\rid$ are the only
attributes, and $U \intersect P = \emptyset$ ensures that constraints
relating $\uid$ and $\rid$ are useless (consequently, any non-zero value
for $w_4$ suffices).  The presence of the ``dummy'' user $u_0$ and
``dummy'' resource $r_0$ ensure that the UAE and RAE in every rule in
$\piabac$ contains a conjunct for $\uid$ or $\rid$, respectively, because
no correct rule can apply to all users or all resources.  These
observations, and the above choice of weights, implies that the WSC of a
rule $\rho$ in $\pirbac$ equals the number of users that satisfy $\rho$
plus the number of resources (i.e., permissions) that satisfy $\rho$.
Thus, $\wscPolL(\pirbac)$ equals the number of edges in the corresponding
RBAC policy, and an ABAC policy with minimum WSC corresponds to an RBAC
policy with minimum number of edges.

\section{Asymptotic Running Time}
\label{sec:aymptotic-time}

This section analyzes the asymptotic running time of our algorithm.  We
first analyze the main loop in Figure \ref{fig:alg}, i.e., the while loop in
lines \ref{tl:while-uncov}--\ref{tl:endwhile-uncov}. First consider the cost
of one iteration.  The running time of $\candidateconstraint$ in line
\ref{tl:cc} is $O(|\Au|\times|\Ar|)$.  The running time of line \ref{tl:s}
is $O(|U_{r,o}|\times|\Au|\times|\Ar|)$, where $U_{r,o} = \{u'\in U \;| \;
\tuple{u',r,o}\in \uncovup\}$; this running time is achieved by
incrementally maintaining an auxiliary map that maps each pair $\tuple{r,o}$
in $R\times\Op$ to $U_{r,o}$.  The running time of function
$\generalizerule$ in line \ref{ccr:generalizerule} in Figure
\ref{fig:addcandidaterule} is $O(|2^{|\cc|}|)$. Other steps in the main loop
are either constant time or linear, i.e., $O(|\Au|+|\Ar|+|\up_0|)$.  Now
consider the number of iterations of the main loop.  The number of
iterations is $|\Rho_1|$, where $\Rho_1$ is the set of rules generated by
the main loop.  In the worst case, the rule generated in each iteration
covers one user-permission tuple, and $|\Rho_1|$ is as large as $|\up_0|$.
Typically, rules generalize to cover many user-permission tuples, and
$|\Rho_1|$ is much smaller than $|\up_0|$.

The running time of function $\mergerules$ is $O(|\Rho_1|^3)$.  The running
time of function $\simplifyrules$ is based on the running times of the five
``elim''
functions that it calls. Let $\lc$ (mnemonic for ``largest conjunct'')
denote the maximum number of sets in a conjunct for a multi-valued user
attribute in the rules in $\Rho_1$, i.e., $\forall a\in\Aum.\, \forall
\rho\in\Rho_1.\, |\uae(\rho)(a)|\leq \lc$.  The value of $\lc$ is at most
$|\valm|$ but typically small (one or a few).  The running time of function
$\elimredundantsets$ is $O(|\Au|\times \lc^2 \times |\vals|)$.  Checking
validity of a rule $\rho$ takes time linear in $|\mean{\rho}|$. Let $\lm$
(mnemonic for ``largest meaning'') denote the maximum value of
$|\mean{\rho}|$ among all rules $\rho$ passed as the first argument in a
call to $\elimconstraints$, $\elimconjuncts$, or $\elimelements$.  The
value of $\lm$ is at most $|\up_0|$ but typically much smaller.  The
running time of function $\elimconstraints$ is $O((2^{|\cc|}) \times \lm)$.
The running time of function $\elimconjuncts$ is $O((2^{|\Au|}+2^{|\Ar|})
\times \lm)$.  The exponential factors in the running time of
$\elimconstraints$ and $\elimconjuncts$ are small in practice, as discussed
above; note that the factor of $\lm$ represents the cost of checking
validity of a rule.  The running time of $\elimelements$ is $O(|\Au|\times
\lm)$.  Let $\lexpr$ (mnemonic for ``largest expressions'') denote the
maximum of $\wsc(\uae(\rho)) + \wsc(\rae(\rho))$ among rules $\rho$
contained in any set $\Rho$ passed as the first argument in a call to
$\simplifyrules$.  The running time of $\elimoverlapVal$ is
$O(|\Rho_1|\times(|\Au| +|\Ar|)\times \lexpr)$.  The running time of
$\elimoverlapOp$ is $O(|\Rho_1|\times |\Op| \times \lexpr)$.  The
factor $\lexpr$ in the running times of $\elimoverlapVal$ and
$\elimoverlapOp$ represents the cost of subset checking.  The
number of iterations of the while loop in line
\ref{tl:simplifyrules}--\ref{tl:endwhile-simplifyrules} is $|\Rho_1|$ in
the worst case.  The overall running time of the algorithm is worst-case
cubic in $|\up_0|$.

\section{Processing Order}
\label{sec:processing-order}

This section describes the order in which tuples and rules are processed by
our algorithm.

When selecting an element of $\uncovup$ in line \ref{tl:select-uncov} of
the top-level pseudocode in Figure \ref{fig:alg}, the algorithm selects
the user-permission tuple with the highest (according to lexicographic
order) value for the following quality metric $\QupL$, which maps
user-permission tuples to triples.  Informally, the first two components of
$\QupL(\tuple{u,r,o})$ are the frequency of permission $p$ and user $u$,
respectively, i.e., their numbers of occurrences in $\up_0$, and the third
component is the string representation of $\tuple{u,r,o}$ (a deterministic
although somewhat arbitrary tie-breaker when the first two components of
the metric are equal).
\begin{eqnarray*}
  \freq(\tuple{r,o}) &=& |\setc{\tuple{u',r',o'}\in\up_0}{r'=r \land o'=o}|\\
  \freq(u) &=& |\setc{\tuple{u',r',o'}\in\up_0}{u'=u}|\\
  \hypertarget{Qup}{\Qup(\tuple{u,r,o})} &=& \tuple{\freq(\tuple{r,o}), \freq(u), \toString(\tuple{u,r,o})}
\end{eqnarray*}


In the iterations over $\Rho$ in $\mergerulesL$ and $\simplifyrulesL$, the
order in which rules are processed is deterministic in our implementation,
because $\Rho$ is implemented as a linked list, loops iterate over the
rules in the order they appear in the list, and newly generated rules are
added at the beginning of the list.  In $\mergerulesL$, the workset is a
priority queue sorted in descending lexicographic order of rule pair
quality, where the quality of a rule pair $\tuple{\rho_1,\rho_2}$ is
$\tuple{\max(\QrulL(\rho_1), \QrulL(\rho_2)), \min(\QrulL(\rho_1),
  \QrulL(\rho_2))}$.

\section{Optimizations}
\label{sec:optimizations}

\myparagraph{Periodic Merging of Rules.}

Our algorithm processes $\up_0$ in batches of 1000 tuples, and calls
$\mergerulesL$ after processing each batch.  Specifically, 1000 tuples are
selected at random from $\uncovup$, they are processed in the order
described in Section \ref{sec:processing-order}, $\mergerulesL(\Rho)$ is
called, and then another batch of tuples is processed.



This heuristic optimization is motivated by the observation that merging
sometimes has the side-effect of generalization, i.e., the merged rule may
cover more tuples than the rules being merged.  Merging earlier (compared
to waiting until $\uncovup$ is empty) allows additional tuples covered by
merged rules to be removed from $\uncovup$ before those tuples are
processed by the loop over $\uncovup$ in the top-level pseudocode in Figure
\ref{fig:alg}.  Without this heuristic optimization, those tuples would be
processed by the loop over $\uncovup$, additional rules would be generated
from them, and those rules would probably later get merged with other
rules, leading to the same policy.


\myparagraph{Caching}

To compute $\mean{\rho}$ for a rule $\rho$, our algorithm first computes
$\mean{\uae(\rho)}$ and $\mean{\rae(\rho)}$.  As an optimization, our
implementation caches $\mean{\rho}$, $\mean{\uae(\rho)}$, and
$\mean{\rae(\rho)}$ for each rule $\rho$.  Each of these values is stored
after the first time it is computed.  Subsequently, when one of these
values is needed, it is recomputed only if some component of $\rho$,
$\uae(\rho)$ or $\rae(\rho)$, respectively, has changed.  In our
experiments, this optimization improves the running time by a factor of
approximately 8 to 10.

\myparagraph{Early Stopping.}

In the algorithm without noise detection, in $\mergerulesL$, when checking
validity of $\rhomerge$, our algorithm does not compute $\mean{\rhomerge}$
completely and then test $\mean{\rhomerge}\subseteq \up_0$.  Instead, as it
computes $\mean{\rhomerge}$, it immediately checks whether each element is
in $\up_0$, and if not, it does not bother to compute the rest of
$\mean{\rhomerge}$.  In the algorithm with noise detection, that validity
test is replaced with the test $|\mean{\rhomerge}\setminus\up_0| \div
|\mean{\rhomerge}| \leq \alpha$.  We incrementally compute the ratio on the
left while computing $\mean{\rhomerge}$, and if the ratio exceeds
$2\alpha$, we stop computing $\mean{\rhomerge}$, under the assumption that
$\rhomerge$ would probably fail the test if we continued.  This heuristic
decreases the running time significantly.  It can affect the result, but it
had no effect on the result for the problem instances on which we evaluated
it.


%
%
%

\section{Details of Sample Policies}
\label{sec:sample-policy-details}

The figures in this section contain all rules and some illustrative
attribute data for each sample policy.  This section also describes in more
detail the manually written attribute datasets and synthetic attribute
datasets for the sample policies.

The policies are written in a concrete syntax with the following kinds of
statements.  userAttrib(${\it uid}, a_1=v_1, a_2=v_2, \ldots$) provides
user attribute data for a user whose ``uid'' attribute equals {\it uid} and
whose attributes $a_1, a_2, \ldots$ equal $v_1, v_2, \ldots$, respectively.
The resourceAttrib statement is similar.  The statement rule({\it uae};
{\it pae}; {\it ops}; {\it con}) defines a rule; the four components of
this statement correspond directly to the four components of a rule as
defined in Section \ref{sec:language}.  In the attribute expressions and
constraints, conjuncts are separated by commas.  In constraints, the
superset relation ``$\supseteq$'' is denoted by ``\textgreater'', the
contains relation ``$\ni$'' is denoted by ``]'', and the
superset-of-an-element-of relation $\supseteqin$ is denoted by
``supseteqIn''.

\myparagraph{University Sample Policy}

In our university sample policy, user attributes include position (applicant,
student, faculty, or staff), department (the user's department), crsTaken
(set of courses taken by a student), crsTaught (set of courses for which
the user is the instructor (if the user is a faculty) or the TA (if the
user is a student), and isChair (true if the user is the chair of his/her
department).  Resource attributes include type (application, gradebook,
roster, or transcript), crs (the course a gradebook or roster is for, for
those resource types), student (the student whose transcript or application
this is, for type=transcript or type=application), and department (the
department the course is in, for $\mbox{type} \in \set{\mbox{gradebook},
  \mbox{roster}}$; the student's major department, for type=transcript).
The policy rules and illustrative userAttrib and resourceAttrib statements
appear in Figure \ref{fig:university}.  The constraint ``crsTaken ]
crs'' in the first rule for gradebooks ensures that a user can apply the
readMyScores operation only to gradebooks for courses the student has
taken.  This is not essential, but it is natural and is advisable according
to the defense-in-depth principle.

\newcommand{\napp}{n_{\rm app}} 
\newcommand{\nstu}{n_{\rm stu}}
\newcommand{\nfac}{n_{\rm fac}} 
\newcommand{\ncrs}{n_{\rm crs}} 

The manually written attribute dataset for this sample policy contains a
few instances of each type of user and resource: two academic departments,
a few faculty, a few gradebooks, several students, a few staff in each of
two administrative departments (admissions office and registrar), etc.  We
generated a series of synthetic attribute datasets, parameterized by the
number of academic departments.  The generated userAttrib and
resourceAttrib statements for the users and resources associated with each
department are similar to but more numerous than the userAttrib and
resourceAttrib statements in the manually written dataset.  For each
department, a scaling factor called the {\it department size} is selected
from a normal distribution with mean 1 and standard deviation 1, truncated
at 0.5 and 5 (allowing a 10-to-1 ratio between the sizes of the largest and
smallest departments).  The numbers of applicants, students, faculty, and
courses associated with a department equal the department size times
$\napp$, $\nstu$, $\nfac$, and $\ncrs$, respectively, where $\napp = 5$,
$\nstu = 20$, $\nfac = 5$, $\ncrs = 10$.  The numbers of courses taught by
faculty, taken by students, and TAd by students are selected from
categorical distributions with approximately normally distributed
probabilities.  The courses taught by faulty are selected uniformly.  The
courses taken and TAd by students are selected following a Zipf
distributions, to reflect the varying popularity of courses.  The number of
staff in each administrative department is proportional to the number of
academic departments.

\begin{figure}[tb]
\begin{tabbing}
// Rules for Gradebooks\\
// A user can read his/her own scores in gradebooks\\
// for courses he/she has taken.\\
rule(; type=gradebook; {readMyScores}; crsTaken ] crs)\\
// A user (the instructor or TA) can add scores and\\
// read scores in the gradebook for courses he/she\\
// is teaching.\\
rule(\=; type=gradebook; \{addScore, readScore\}; \\
\> crsTaught ] crs;)\\
// The instructor for a course (i.e., a faculty teaching\\
// the course) can change scores and assign grades in \\
// the gradebook for that course.\\
rule(\=position=faculty; type=gradebook; \\
\> \{changeScore, assignGrade\}; crsTaught ] crs)\\
\\
// Rules for Rosters\\
// A user in registrar's office can read and modify all\\
// rosters.\\
rule(department=registrar; type=roster; \{read, write\}; )\\
// The instructor for a course (i.e., a faculty teaching\\
// the course) can read the course roster.\\
rule(\=position=faculty; type=roster; \{read\};\\
\> crsTaught ] crs)\\
\\
// Rules for Transcripts\\
// A user can read his/her own transcript.\\
rule(; type=transcript; \{read\}; uid=student)\\
// The chair of a department can read the transcripts\\
// of all students in that department.\\
rule(\=isChair=true; type=transcript; \{read\}; \\
\> department=department)\\
// A user in the registrar's office can read every\\
// student's transcript.\\
rule(department=registrar; type=transcript; \{read\}; )\\
\\
// Rules for Applications for Admission\\
// A user can check the status of his/her own application.\\
rule(; type=application; \{checkStatus\}; uid=student)\\
\\
// A user in the admissions office can read, and\\
// update the status of, every application.\\
rule(\=department=admissions; type=application;\\
\> \{read, setStatus\}; )\\
\\
// An illustrative user attribute statement.\\
userAttrib(\=csFac2, position=faculty, department=cs, \\
\> crsTaught=\{cs601\})\\
// An illustrative resource attribute statement.\\
resourceAttrib(\=cs601gradebook, department=cs,\\
\> crs=cs601, type=gradebook)
\end{tabbing}
\caption{University sample policy.}
\label{fig:university}
\end{figure}

\myparagraph{Health Care Sample Policy}

In our health care sample policy, user attributes include position (doctor or
nurse; for other users, this attribute equals $\bot$), specialties (the
medical areas that a doctor specializes in), teams (the medical teams a
doctor is a member of), ward (the ward a nurse works in or a patient is
being treated in), and agentFor (the patients for which a user is an
agent).  Resource attributes include type (HR for a health record, or
HRitem for a health record item), patient (the patient that the HR or HR
item is for), treatingTeam (the medical team treating the aforementioned
patient), ward (the ward in which the aforementioned patient is being
treated), author (author of the HR item, for type=HRitem), and topics
(medical areas to which the HR item is relevant, for type=HRitem).  The
policy rules and illustrative userAttrib and resourceAttrib statements
appear in Figure \ref{fig:healthcare}.

\newcommand{\npat}{n_{\rm pat}}
\newcommand{\nnurse}{n_{\rm nurse}}
\newcommand{\ndoc}{n_{\rm doc}}
\newcommand{\nagent}{n_{\rm ag}}
\newcommand{\nteam}{n_{\rm team}}

The manually written attribute dataset for this sample policy contains a
small number of instances of each type of user and resource: a few nurses,
doctors, patients, and agents, two wards, and a few items in each patient's
health record.  We generated a series of synthetic attribute datasets,
parameterized by the number of wards.  The generated userAttrib and
resourceAttrib statements for the users and resources associated with each
ward are similar to but more numerous than the userAttrib and
resourceAttrib statements in the manually written dataset.  For each ward,
a scaling factor called the {\it ward size} is selected from a normal
distribution with the same parameters as the department size distribution
described above.  The numbers of patients, nurses, doctors, agents, and
teams associated with a ward equal the ward size times $\npat$, $\nnurse$,
$\ndoc$, $\nagent$, and $\nteam$, respectively, where $\npat=10$,
$\ndoc=2$, $\nnurse=4$, $\nagent=2$, and $\nteam=2$.  The numbers of items
in each patient's medical record, the topics associated with each HR item,
patients associated with each agent, specialties per doctor, and teams each
doctor is on are selected from categorical distributions with approximately
normally distributed probabilities.  The topics and specialties associated
with HR items and doctors, respectively, are selected following a Zipf
distribution.

\begin{figure}[tb]
\begin{tabbing}
// Rules for Health Records\\
// A nurse can add an item in a HR for a patient in\\
// the ward in which he/she works.\\
rule(position=nurse; type=HR; \{addItem\}; ward=ward)\\
// A user can add an item in a HR for a patient treated\\
// by one of the teams of which he/she is a member.\\
rule(; type=HR; \{addItem\}; teams ] treatingTeam)\\
// A user can add an item with topic "note" in his/her\\
// own HR.\\
rule(; type=HR; \{addNote\}; uid=patient)\\
// A user can add an item with topic "note" in the HR\\
// of a patient for which he/she is an agent.\\
rule(; type=HR; \{addNote\}; agentFor ] patient)\\
\\
// Rules for Health Record Items\\
// The author of an item can read it.\\
rule(; type=HRitem; \{read\}; uid=author)\\
// A nurse can read an item with topic "nursing" in a HR\\
// for a patient in the ward in which he/she works.\\
rule(\=position=nurse; type=HRitem, \\
\> topics supseteqIn \{\{nursing\}\}; \{read\}; ward=ward)\\
// A user can read an item in a HR for a patient treated\\
// by one of the teams of which he/she is a member, if \\
// the topics of the item are among his/her specialties.\\
rule(\=; type=HRitem; \{read\}; specialties \textgreater\ topics,\\
\> teams ] treatingTeam)\\
// A user can read an item with topic "note" in his/her\\
// own HR.\\
rule(\=; type=HRitem, topics supseteqIn \{\{note\}\}; \{read\};\\
\> uid=patient)\\
// An agent can read an item with topic "note" in the\\
// HR of a patient for which he/she is an agent.\\
rule(\=; type=HRitem, topics supseteqIn \{\{note\}\}; \{read\}; \\
\>agentFor ] patient)\\
// An illustrative user attribute statement.\\
userAttrib(\=oncDoc1, position=doctor, \\
\> specialties=\{oncology\}, \\
\> teams=\{oncTeam1, oncTeam2\})\\
// An illustrative resource attribute statement.\\
resourceAttrib(\=oncPat1nursingItem, type=HRitem, \\
\> author=oncNurse2, patient=oncPat1, \\
\> topics=\{nursing\}, ward=oncWard,\\
\> treatingTeam=oncTeam1)
\end{tabbing}
\caption{Health care sample policy.}
\label{fig:healthcare}
\end{figure}

\myparagraph{Project Management Sample Policy}

In our project management sample policy, user attributes include projects
(projects the user is working on), projectsLed (projects led by the user),
adminRoles (the user's administrative roles, e.g., accountant, auditor,
planner, manager), expertise (the user's areas of technical expertise, e.g.,
design, coding), tasks (tasks assigned to the user), department (department
that the user is in), and isEmployee (true if the user is an employee, false
if the user is a contractor).  Resource attributes include type (task,
schedule, or budget), project (project that the task, schedule, or budget is
for), department (department that the aforementioned project is in),
expertise (areas of technical expertise required to work on the task, for
type=task) and proprietary (true if the task involves proprietary
information, which is accessible only to employees, not contractors).
The policy rules and illustrative userAttrib and resourceAttrib statements
appear in Figure \ref{fig:project-management}.

\newcommand{\nproj}{n_{\rm proj}}
\newcommand{\nnemp}{n_{\rm nonEmp}}
\newcommand{\nemp}{n_{\rm emp}}

The manually written attribute dataset for this sample policy contains a
small number of instances of each type of user (managers, accountants,
coders, non-employees (e.g,, contractors) and employees with various areas
of expertise, etc.) and each type of resource (two departments, two
projects per department, three tasks per project, etc.).  We generated a
series of synthetic attribute datasets, parameterized by the number of
departments.  The generated userAttrib and resourceAttrib statements for
the users and resources associated with each department are similar to but
more numerous than the userAttrib and resourceAttrib statements in the
manually written dataset.  For each department, a scaling factor called the
{\it department size} is selected from a normal distribution with the same
parameters as for the university sample policy.  The numbers of projects,
non-employees per area of expertise, and employees per area of expertise
that are associated with a department equal the department size times
$\nproj$, $\nnemp$, and $\nemp$, respectively, where $\nproj=2$,
$\nnemp=1$, and $\nemp=1$.  
Each department has one manager, accountant, auditor, and planner, and one
project leader per project.  Six tasks are associated with each project.
The projects assigned to non-employees and employees are selected following
a Zipf distribution.



\begin{figure}[tb]
\begin{tabbing}
// The manager of a department can read and approve\\
// the budget for a project in the department.\\
rule(adminRoles supseteqIn \{\{manager\}\}; type=budget;\\
~~~~~~\= \{read approve\}; department=department)\\
// A project leader can read and write the project\\
// schedule and budget.\\
rule( ; type in \{schedule, budget\}; \{read, write\};\\
\> projectsLed ] project)\\
// A user working on a project can read the project\\
// schedule.\\
rule( ; type=schedule; \{read\}; projects ] project)\\
// A user can update the status of tasks assigned to\\
// him/her.\\
rule( ; type=task; \{setStatus\}; tasks ] rid)\\
// A user working on a project can read and request\\
// to work on a non-proprietary task whose required\\
// areas of expertise are among his/her areas of\\
// expertise.\\
rule( ; type=task, proprietary=false; \{read request\}; \\
\> projects ] project, expertise \textgreater\ expertise)\\
// An employee working on a project can read and\\
// request to work on any task whose required areas\\
// of expertise are among his/her areas of expertise.\\
rule(isEmployee=True; type=task; \{read request\}; \\
\> projects ] project, expertise \textgreater\ expertise)\\
// An auditor assigned to a project can read the\\
// budget.\\
rule(adminRoles supseteqIn \{\{auditor\}\}; type=budget;\\
\> \{read\}; projects ] project)\\
// An accountant assigned to a project can read and\\
// write the budget.\\
rule(adminRoles supseteqIn \{\{accountant\}\};\\
\> type=budget; \{read, write\}; projects ] project)\\
// An accountant assigned to a project can update the\\
// cost of tasks.\\
rule(adminRoles supseteqIn \{\{accountant\}\}; type=task; \\
\> \{setCost\}; projects ] project)\\
// A planner assigned to a project can update the\\
// schedule.\\
rule(adminRoles supseteqIn \{\{planner\}\};\\
\> type=schedule; \{write\}; projects ] project)\\
// A planner assigned to a project can update the\\
// schedule (e.g., start date, end date) of tasks.\\
rule(adminRoles supseteqIn \{\{planner\}\}; type=task; \\
\> \{setSchedule\}; projects ] project)\\
// An illustrative user attribute statement.\\
userAttrib(des11, expertise=\{design\}, projects=\{proj11\}, \\
\> isEmployee=True,\\
\> tasks=\{proj11task1a, proj11task1propa\})\\
// An illustrative resource attribute statement.\\
resourceAttrib(proj11task1a, type=task, project=proj11,\\
\> department=dept1, expertise=\{design\},\\
\> proprietary=false)
\end{tabbing}
\caption{Project management sample policy.}
\label{fig:project-management}
\end{figure}

\section{Example: Processing of a user-permission tuple}
\label{sec:working-example}

\begin{figure*}[htb]
  \centering
  \includegraphics[width=180mm]{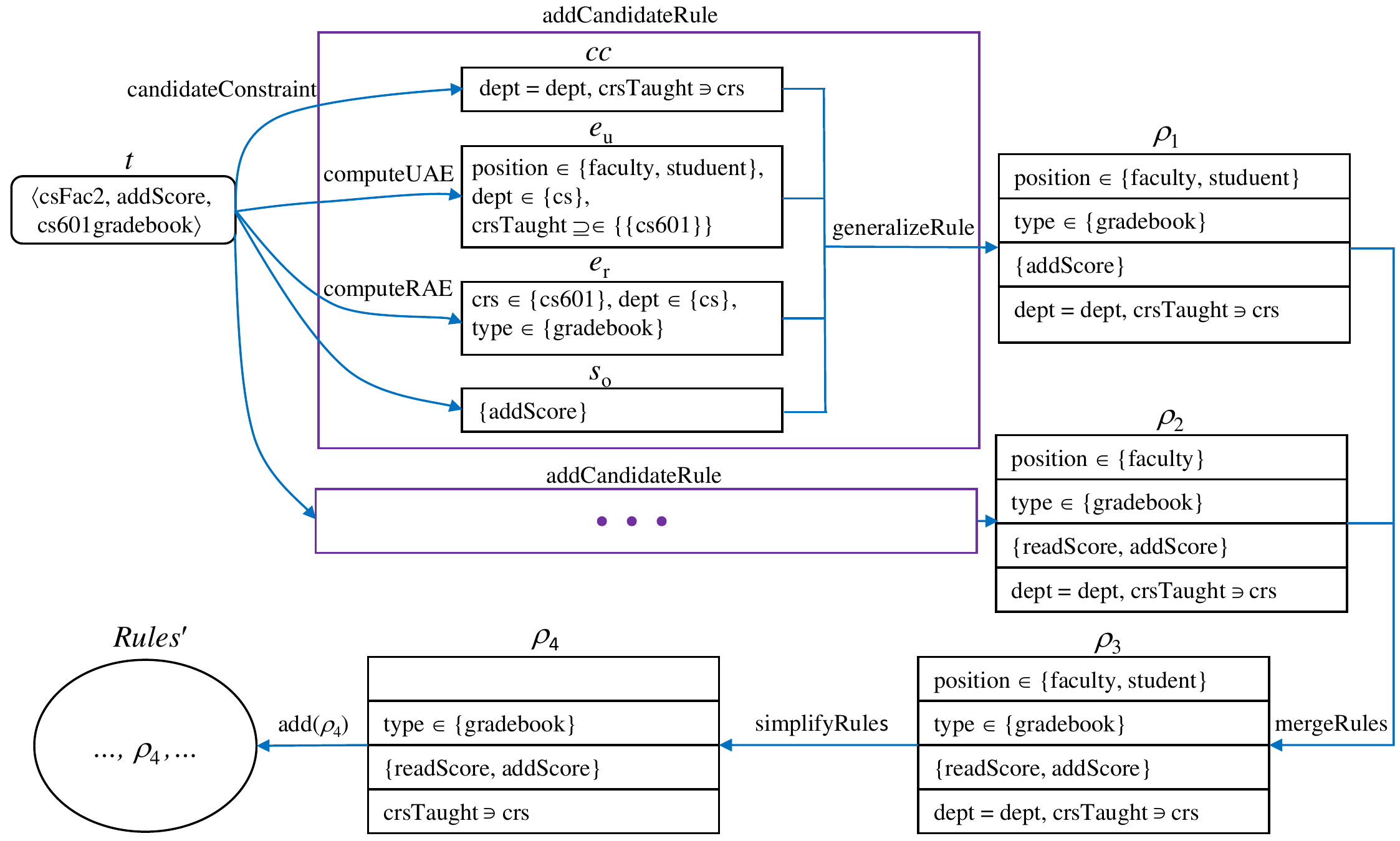}
  \caption{Diagram representing the processing of one user-permission tuple
    selected as a seed, in the university sample policy.  Rules are depicted
    as rectangles with four compartments, corresponding to the four
    components of a rule tuple.}
  \label{fig:working-example}
\end{figure*}

Figure \ref{fig:working-example} illustrates the processing of the
user-permission tuple $t = \tuple{{\rm csFac2}, {\rm addScore}, {\rm
    cs601gradebook}}$ selected as a seed (i.e., selected in line
\ref{tl:select-uncov} of Figure \ref{fig:alg}), in a smaller version of
the university sample policy containing only one rule, namely, the second rule
in Figure \ref{fig:university}.
Attribute data for user {\rm csFac2} and resource {\rm cs601gradebook}
appear in Figure \ref{fig:university}.

The edge from $t$ to $cc$ labeled ``$\candidateconstraint$'' represents the
call to $\candidateconstraintL$, which returns the set of atomic
constraints that hold between csFac2 and cs601gradebook; these constraints
are shown in the box labeled $\cc$.  The two boxes labeled
``$\addcandidaterule$'' represent the two calls to $\addcandidateruleL$.
Internal details are shown for the first call but elided for the second
call.  The edges from $t$ to $\eu$ and from $t$ to $\er$ represent the
calls in $\addcandidateruleL$ to $\computeuae$ and $\computerae$,
respectively.  The call to $\computeuae$ returns a user-attribute
expression $\eu$ that characterizes the set $\su$ containing users $u'$
with permission $\tuple{{\rm addScore}, {\rm cs601gradebook}}$ and such
that $\candidateconstraint({\rm cs601gradebook}, u')=\cc$.  The call to
$\computerae$ returns a resource-attribute expression that characterizes
$\set{{\rm cs601gradebook}}$.  The set of operations considered in this
call to $\addcandidateruleL$ is simply $\so = \set{{\rm addScore}}$.  The
call to $\generalizerule$ generates a candidate rule $\rho_1$ by assigning
$\eu$, $\er$ and $\so$ to the first three components of $\rho_1$, and
adding the two atomic constraints in $\cc$ to $\rho_1$ and eliminating the
conjuncts in $\eu$ and $\er$ corresponding to the attributes mentioned in
$\cc$.  Similarly, the second call to $\addcandidateruleL$ generates
another candidate rule $\rho_2$.  The call to $\mergerules$ merges $\rho_1$
and $\rho_2$ to form $\rho_3$,
which is simplified by the call to $\simplifyrulesL$ to produce a
simplified rule $\rho_4$, which is added to candidate rule set $\Rho'$.

\section{Syntactic Similarity}
\label{sec:syntactic-similarity}

Syntactic similarity of policies measures the syntactic similarity of rules
in the policies. The {\em syntactic similarity} of rules $\rho$ and $\rho'$,
denoted $\synSim(\rho,\rho')$, is defined by
\begin{eqnarray*}
  \uaeSim(e,e') &=& |\Au|^{-1}\sum_{a\in\Au} J(e(a), e'(a))\\
  \raeSim(e,e') &=& |\Ar|^{-1}\sum_{a\in\Ar} J(e(a), e'(a))\\
  \synSim(\rho,\rho') &=& \average(
  \begin{array}[t]{@{}l@{}}
    \uaeSim(\uae(\rho),\uae(\rho')),
    \raeSim(\rae(\rho),\rae(\rho')),\\
    J(\ops(\rho),\ops(\rho')),\, J(\con(\rho),\con(\rho')))
  \end{array}
\end{eqnarray*}
where the Jaccard similarity of two sets is $J(S_1, S_2)=|S_1\intersect S_2| /
|S_1 \union S_2|$.

The {\em syntactic similarity} of rule sets $\Rho$ and $\Rho'$ is the
average, over rules $\rho$ in $\Rho$, of the syntactic similarity between
$\rho$ and the most similar rule in $\Rho'$.  The {\em syntactic
  similarity} of policies is the maximum of the syntactic similarity of the
sets of rules in the policies, considered in both orders (this makes the
relation symmetric).
\begin{eqnarray*}
  \synSim(\Rho,\Rho') &=& 
  \begin{array}[t]{@{}l@{}}
    |\Rho|^{-1} \times{} \\
    \sum_{\rho\in\Rho}\max(\setc{\synSim(\rho,\rho')}{\rho'\in \Rho'}))
  \end{array}\\
  \synSim(\pi,\pi') &=& \max(
  \begin{array}[t]{@{}l@{}}
    \synSim(\rules(\pi), \rules(\pi')),\\
    \synSim(\rules(\pi'), \rules(\pi)))
  \end{array}
\end{eqnarray*}

\section{ROC Curves for Noise Detection Parameters}
\label{sec:ROC}

\begin{figure}[htb]
  \centering
  \includegraphics[width=85mm]{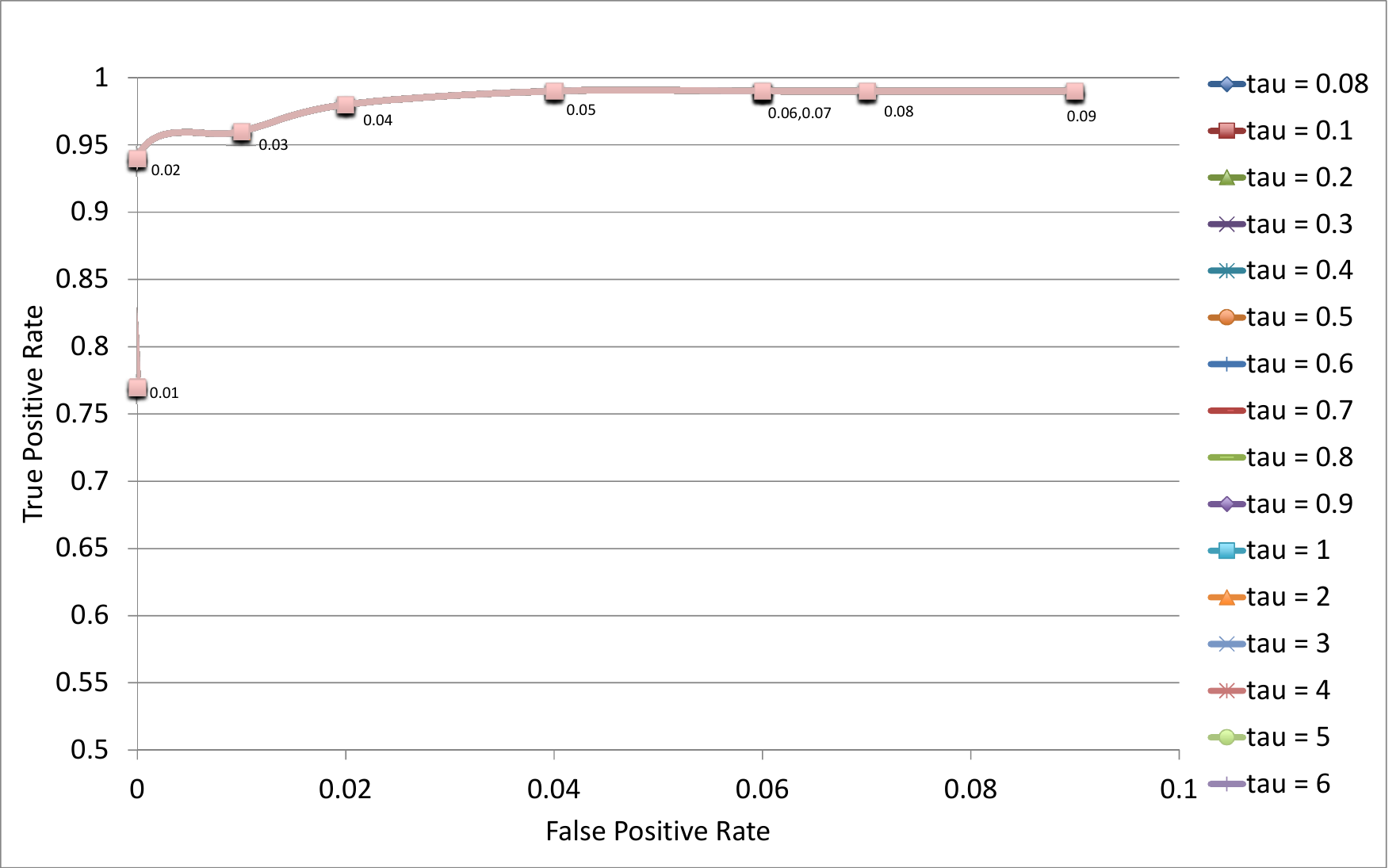}
  \caption{ROC curve showing shows the dependence of the true positive rate
    (TPR) and false positive rate (FPR) for under-assignments on $\alpha$
    and $\tau$.}
  \label{fig:ROC-under-assignments}
\end{figure}

\begin{figure}[htb]
  \centering
  \includegraphics[width=85mm]{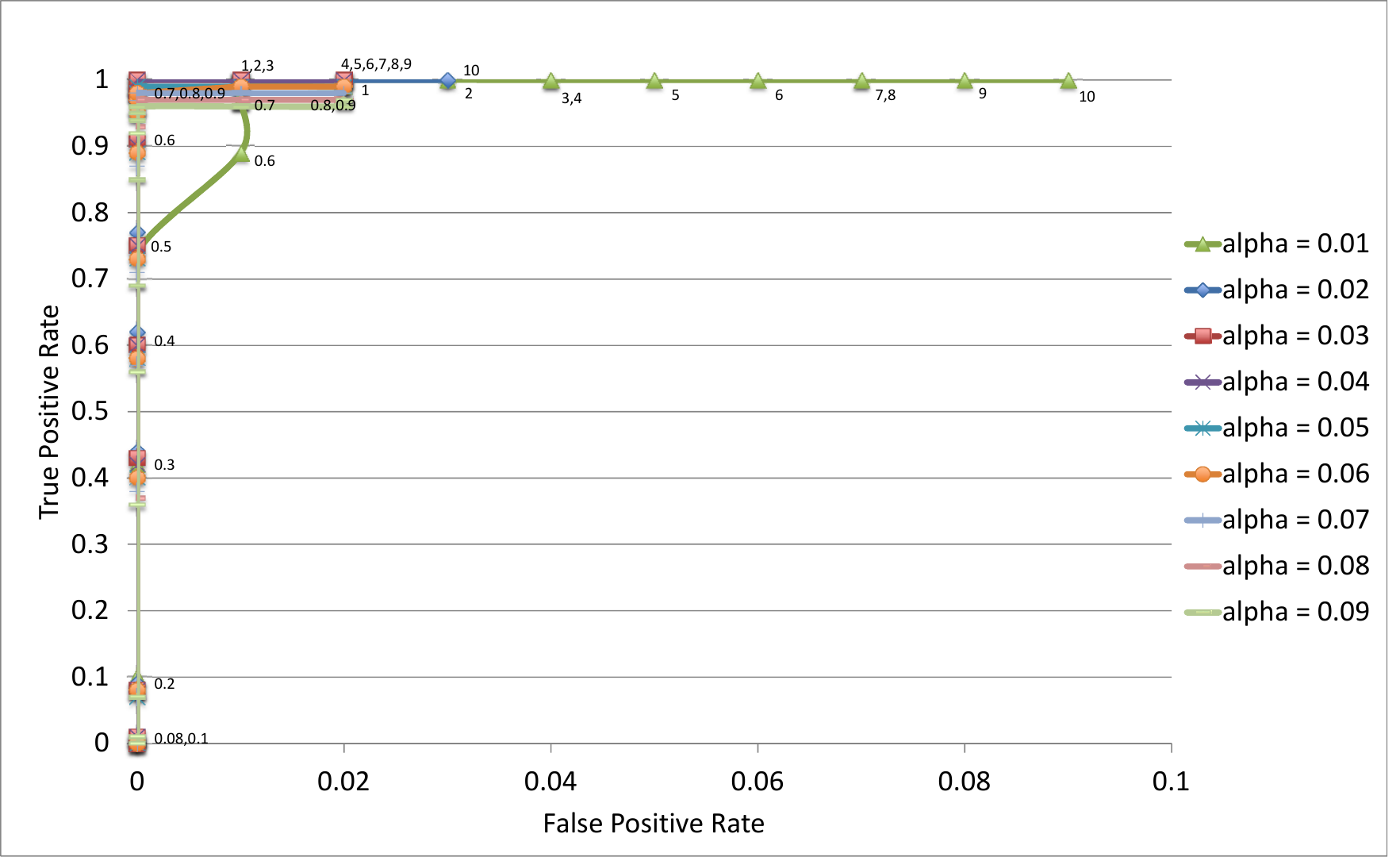}
  \caption{ROC curve showing shows the dependence of the true positive rate
    (TPR) and false positive rate (FPR) for over-assignments on $\alpha$
    and $\tau$.}
  \label{fig:ROC-over-assignments}
\end{figure}

When tuning the parameters $\alpha$ and $\tau$ used in noise detection (see
Section \ref{sec:algorithm:noise}), there is a trade-off between true
positives and false positives.  To illustrate the trade-off, the Receiver
Operating Characteristic (ROC) curve in Figure
\ref{fig:ROC-under-assignments} shows the dependence of the true positive
rate (TPR) and false positive rate (FPR) for under-assignments on $\alpha$
and $\tau$ for synthetic policies with 20 rules and 6\% noise, split
between under-assignments and over-assignments as described in Section
\ref{sec:eval:noise}.  Figure \ref{fig:ROC-over-assignments} shows the TPR
and FPR for over-assignments.  Each data point is an average over 10
synthetic policies.  In each of these two sets of experiments, true
positives are reported noise (of the specified type, i.e., over-assignments
or under-assignments) or that are also actual noise; false negatives are
actual noise that are not reported; false positives are reported noise that
are not actual noise; and true negatives are user-permission tuples that
are not actual noise and are not reported as noise.

Generally, we can see from the ROC curves that, with appropriate parameter
values, it is possible to achieve very high TPR and FPR simultaneously, so
there is not a significant inherent trade-off between them.

From the ROC curve for under-assignments, we see that the value of $\tau$
does not affect computation of under-assignments, as expected, because
detection of under-assignments is performed before detection of
over-assignments (the former is done when each rule is generated, and the
latter is done at the end).  We see from the diagonal portion of the curve
in the upper left that, when choosing the value of $\alpha$, there is a
trade-off between the TPR and FPR, i.e., having a few false negatives and a
few false positives.

From the ROC curve for over-assignments, we see that the value of $\alpha$
affects the rules that are generated, and hence it affects the computation
of over-assignments based on those rules at the end of the rule generation
process.  For $\alpha=0.01$, when choosing $\tau$, there is some trade-off
between the TPR and FPR.  For $\alpha \ge 0.02$, the FPR equals 0
independent of $\tau$, so there is no trade-off: the best values of $\tau$
are the ones with the highest TPR.

\section{Graphs of Results from Experiments with Permission Noise and Attribute Noise}
\label{sec:attrib-noise-graphs}

\begin{figure}[tb]
  \centering
  \includegraphics[width=85mm]{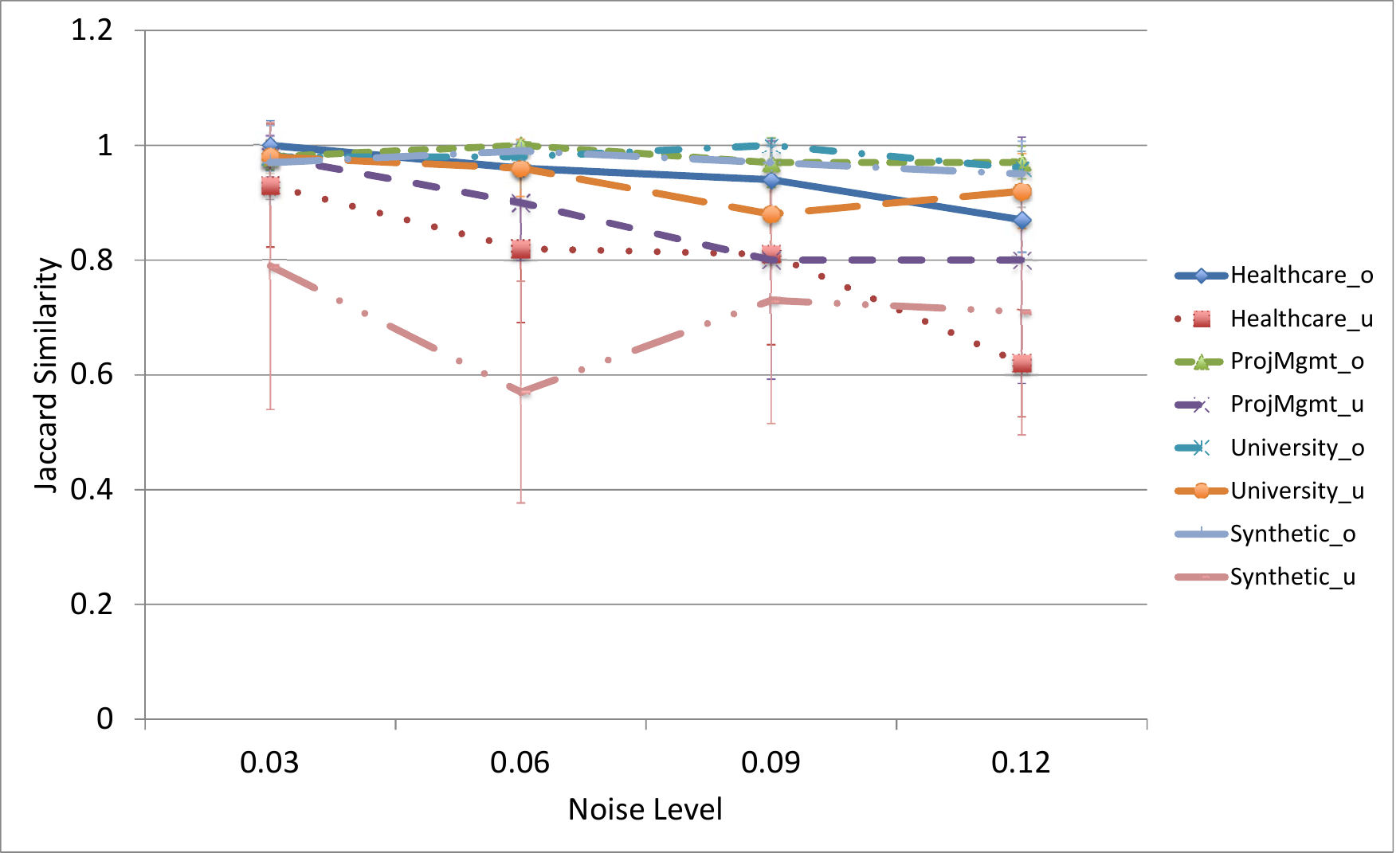}
  \caption{Jaccard similarity of actual and reported under-assignments, and
    Jaccard similarity of actual and reported over-assignments, as a function of
    permission noise level due to permission noise and attribute noise. 
  }
  \label{fig:jaccard-similarity-attrib-noise}
\end{figure}

\begin{figure}[tb]
  \centering
  \includegraphics[width=85mm]{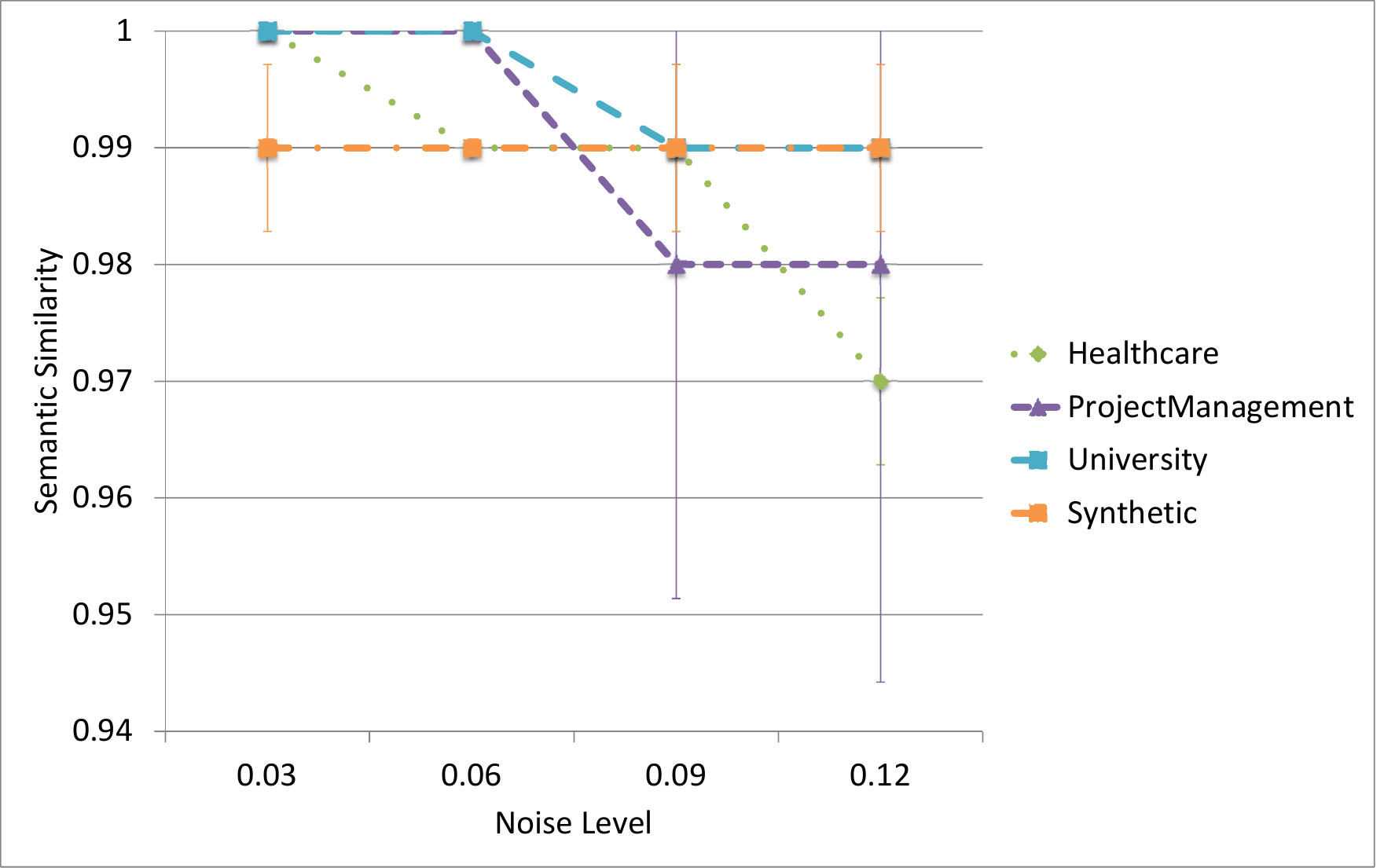}
  \caption{Semantic similarity of the original policy and the mined policy,
    as a function of permission noise level due to permission noise and
    attribute noise.}
  \label{fig:semantic-similarity-attrib-noise}
\end{figure}

For the experiments with permission noise and attribute noise described in
Section \ref{sec:eval:noise}, Figure \ref{fig:jaccard-similarity-attrib-noise}
shows the Jaccard similarity of the actual and reported over-assignments
and the Jaccard similarity of the actual and reported under-assignments,
and Figure \ref{fig:semantic-similarity-attrib-noise} shows the semantic
similarity of the original and mined policies.  Each data point is an
average over 10 policies.  Error bars show 95\% confidence intervals using
Student's t-distribution.

\section{Translation to Inductive Logic Programming}
\label{sec:translation-to-ILP}

This section describes our translation from the ABAC policy mining problem
to inductive logic programming (ILP) as embodied in Progol
\cite{muggleton00theory,muggleton01cprogol}.  Given an ACL policy and
attribute data, we generate a Progol input file, which contains type
definitions, mode declarations, background knowledge, and examples.

\myparagraph{Type Declarations}

Type definitions define categories of objects.  The types {\tt user}, {\tt
  resource}, {\tt operation}, and {\tt attribValAtomic} (corresponding to
$\vals$) are defined by a statement for each constant of that type; for
example, for each user $u$, we generate the statement {\tt user($u$)}.  The
type {\tt attribValSet} (corresponding to $\valm$) is defined by the rules
\begin{verbatim}
attribValSet([]).
attribValSet([V|Vs]) :- attribValAtomic(V),
                        attribValSet(Vs).
\end{verbatim}
For each attribute $a$, we define a type containing the constants that
appear in values of that attribute in the attribute data; for example, for
each value $d$ of the ``department'' attribute, we generate the statement
{\tt departmentType($d$)}.


\myparagraph{Mode Declarations}

Mode declarations restrict the form of rules that Progol considers, by
limiting how each predicate may be used in learned rules.  Each head mode
declaration {\tt modeh($\ldots$)} describes a way in which a predicate may
be used in the head (conclusion) of a learned rule.  Each body mode
declaration {\tt modeb($\ldots$)} describes a way in which a predicate may
be used in the body (premises) of a learned rule.  Each mode declaration
has two arguments.  The second argument specifies, for each argument $a$ of
the predicate, the type of $a$ and whether $a$ may be instantiated with an
input variable (indicated by ``{\tt +}''), an output variable (indicated by
``{\tt -}''), or a constant (indicated by ``{\tt \#}'').  The first
argument, called the {\it recall}, is an integer or {\tt *}, which bounds
the number of values of the output arguments for which the predicate can
hold for given values of the input arguments and constant arguments; ``{\tt
  *}'' indicates no bound.
The specification of predicate arguments as inputs and outputs also limits
how variables may appear in learned rules.
In a learned rule $h$ {\tt :-} $b_1, \ldots, b_n$, every variable of input
type in each premise $b_i$ must appear either with input type in $h$ or
with output type in some premise $b_j$ with $j<i$.

We generate only one head mode declaration:
\begin{verbatim}
 modeh(1, up(+user, +resource, #operation))
\end{verbatim}
This tells Progol to learn rules that define the user-permission predicate
{\tt up}.

For each single-valued user attribute $a$, we generate a body mode
declaration {\tt modeb(1, $a$U(+user, \#$a$Type))}.  For example, the mode
declaration for a user attribute named ``department'' is {\tt modeb(1,
  departmentU(+user, \#departmentType))}.  We append ``{\tt U}'' to the
attribute name to prevent naming conflicts in case there is a resource
attribute with the same name.  Mode declarations for multi-valued user
attributes are defined similarly, except with ``{\tt *}'' instead of {\tt
  1} as the recall.  Mode declarations for resource attributes are defined
similarly, except with {\tt R} instead of {\tt U} appended to the attribute
name.  We tried a variant translation in which we generated a second body
mode declaration for each attribute, using {\tt -$a$Type} instead of {\tt
  \#$a$Type}, but this led to worse results.

We also generate mode declarations for predicates used to express
constraints.  For each single-valued user attribute $a$ and single-valued
resource attribute $\bar a$, we generate a mode declaration {\tt modeb(1,
  $a$U\_equals\_$\bar a$R(+user,+resource))}; the predicate
$a$U\_equals\_$\bar a$R is used to express atomic constraints of the form
$a=\bar a$.  The mode declarations for the predicates used to express the
other two forms of atomic constraints are similar, using user and resource
attributes with appropriate cardinality, and with ``{\tt contains}'' (for
$\ni$) or ``{\tt superset}'' (for $\supseteq$) instead of ``{\tt equals}''
in the name of the predicate.

\myparagraph{Background Knowledge}

The attribute data is expressed as background knowledge.  For each user $u$
and each single-valued user attribute $a$, we generate a statement {\tt
  $a$U($u, v$)} where $v=\uad(u, a)$.  For each user $u$ and each
multi-valued user attribute $a$, we generate a statement {\tt $a$U($u, v$)}
for each $v\in \uad(u, a)$. Background knowledge statements for resource
attribute data are defined similarly.

Definitions of the predicates used to express constraints are also included
in the background knowledge.  For each equality predicate
$a$\_equals\_$\bar a$ mentioned in the mode declarations, we generate a
statement {\tt $a$U\_equals\_$\bar a$R(U,R) :- $a$U(U,X), $\bar a$R(R,X)}.
The definitions of the predicates used to express the other two forms of
constraints are
\begin{tabbing}
  {\tt $a$U\_contains\_$\bar a$R(U,R) :- $a$U(U,X), $\bar a$R(R,X)}.\\
  {\tt $a$U\_superset\_$\bar a$R(U,R) :- }\={\tt setof(X,\,$a$U(U,X),\,SU),}\\
  \> {\tt setof(Y,\,$\bar a$R(R,Y),\,SR),}\\
  \> {\tt superset(SU,SR),}\\
  \> {\tt not(SR==[])}.\\
  {\tt superset(Y,[A|X]) :- }\={\tt element(A,Y),}\\
  \> {\tt superset(Y,X).}\\
  {\tt superset(Y,[]).}
\end{tabbing}
The premise {\tt not(SR==[])} in the definition of {\tt
  $a$U\_superset\_$\bar a$R} is needed to handle cases where the value of
$\bar a$ is $\bot$.  The predicates {\tt setof} and {\tt element} are
built-in predicates in Progol.

\myparagraph{Examples}

A {\it positive example} is an instantiation of a predicate to be learned
for which the predicate holds.  A {\it negative example} is an
instantiation of a predicate to be learned for which the predicate does not
hold.  For each $\tuple{u,r,o}\in U\times R\times \Op$, if
$\tuple{u,r,o}\in \up_0$, then we generate a positive example {\tt
  up($u,r,o$)}, otherwise we generate a negative example {\tt :-
  up($u,r,o$)} (the leading ``{\tt :-}'' indicates that the example is
negative).  The negative examples are necessary because, without them,
Progol may produce rules that hold for instantiations of {\tt up} not
mentioned in the positive examples.



\end{document}
